\documentclass[fleqn]{article}

\newcommand{\titel}
{Bitcoin and Beyond: Exclusively Informational Money}

\usepackage{stmaryrd,amssymb,amsmath,amsthm,url}

\theoremstyle{definition}

\title{\titel}

\author{
	Jan A.\ Bergstra \& Karl de Leeuw
	 \\
\\
  {\small
	  Informatics Institute,
	  University of Amsterdam}\\
	{\small Email: \url{j.a.bergstra@uva.nl}, \url{karl.de.leeuw@xs4all.nl}
	}
}
\date{}

\begin{document}
\maketitle

\begin{abstract}
The famous new money Bitcoin is classified as a technical
informational money (TIM). Besides introducing the idea of a TIM, a more
extreme notion of informational money will be developed: 
exclusively informational money (EXIM).

The informational coins (INCOs) of an EXIM can be in control of an agent
but are not owned by any agent. INCOs of an EXIM cannot be stolen, but they can be lost,
or thrown away.
The difference between an EXIM and a TIM shows up when considering a user
perspective on security matters. Security for an EXIM user is discussed in substantial detail, with 
the remarkable conclusion that computer security (security models, access control, user names,
passwords, firewalls etc.) 
is not always essential for an EXIM, while the application of cryptography based information 
security is unavoidable for the use of an EXIM.

Bitcoin seems to meet the criteria of an EXIM, but the assertion that ``Bitcoin 
is an EXIM'', might also be considered problematic.  As a thought 
experiment we will contemplate Bitguilder, a hypothetical copy of Bitcoin,
cast as an EXIM and its equally hypothetical extension BitguilderPlus. 

A business ethics assessment of Bitcoin is made which reveals a number
of worries. By combining Bitguilder with a so-called technical informational near-money
(TINM) a dual money system, having two units with a fluctuating rate, may be obtained. It seems that
a dual money can remedy some, but not all, of the ethical worries
that arise when contemplating Bitcoin after hypothetically having become a dominant form of money.

The contributions that Bitcoin's designers can potentially make to the evolution of 
EXIMs and TIMs is analyzed
in terms of the update of the portfolio of money related natural kinds that comes with Bitcoin.
\\[5mm]
\end{abstract}

\newpage{\small\tableofcontents}\newpage

\section{Introduction}
\label{intro}
The emergence and the initial survival of the Bitcoin\footnote{%
Disclosure: using the terminology of our paper at the time of writing the 
first author (thinks of himself that he)  ``owns'' 3.40050000 BTC (which were acquired for about 100 Euro
in total 
including bank transfer cost), as an indirect user of three participation services 
provided by agents who seem to be regular victims of cyberattacks. Two of these particular
providers require
disclosure of personal identification data for the indirect users that they are supporting and two of them indicate 
ethical standards constraining which actions an indirect user requests the service to effectuate on their behalf. 
At the time of
writing the revised version two of these participation services are (temporarily so they say) 
closed because of a combination of security problems as well as legal complications.}
open source peer-to-peer network demonstrates that, given today's computer and network technology,
the development of a competitive money-like system of value exchange, storage, and measurement, 
can be achieved trhough mere software engineering, including its underlying logic, mathematics and
computer science. We believe that this fact will stand for quite some time even if eventually
Bitcoin must give way to another yet more sophisticated informational money.

In this paper\footnote{%
This is the second version of the paper. Compared with \texttt{v1} (\url{arXiv:1304.4758v1}) 
the following modifications were made:
in the title monies was replaced by money because the need to choose between ``monies'' and ``moneys'' as 
different plurals is unfortunate. Some 100 typo's and textual issues were taken care of. 
Reference \cite{Bergstra2013a} was included together with some remarks about formalbitcoins 
and about virtual monies based on that reference. The name Icoin was replaced by Cresent-Star-Bitmoney. In
appendix B emphasis was moved to money of account instead of money of measurement and valuation.}
 we will experiment with the speculative notion of an exclusively informational money (EXIM),
and we will perform a thought experiment with Bitguilder, a hypothetical copy of Bitcoin merely seen as a piece
of software engineering. Bitguilder is uncommitted to a part of the ideological basis that Bitcoin 
supporters may intend or need to keep in reserve for Bitcoin. Bitguilder is classified as an EXIM.

In~\cite{Bergstra2012d} the use of the 
term informaticology has been advocated. In that paper informaticology (IY) is decomposed as: 
IY = CS + DS + FS = Computer Science + Data Science + Fiction Science. Informational money is rooted
in each of these components: CS for protocols, encryption, computing, security, networking; DS for
analysis on distributed data sets resulting from transaction logging; FS for the computer game outlook
that successful informational monies must probably display. This paper is intended to constitute 
a contribution to the informaticology of informational money.

\subsection{Gesell, Maududi, and Nakamoto}
While Bitcoin seems to have been designed so as to be completely international and with some rigorous 
built in protection against inflation, the vision that it incorporates appears to stand at the opposite end of the 
vision on free money that was promoted by Silvio Gesell (1862--1930) who approved of
regional monies which have progressive inflation rigorously built in.\footnote{%
This mechanism was called demurrage, it reappeared later as stamped money. A Bitcoin-like system named Freicoin
 that includes demurrage as one of its features has become operational, (see \cite{Steadman2013}).}
The advantage of enhanced inflation for a regional money is supposed to be that it helps to promote local spending. 
In modern terminology, the mechanism works against globalization. 
A century later, Nakamoto appears in the role of an economic counterpart to Gesell. Gesell may be seen as an icon 
of the movement of regional monies and local exchange and trade systems which are introduced and maintained 
throughout the world often out of idealistic intentions. Nakamoto's idea is equally principled with regards to
inflation as Gesell's views were, though with an opposite outcome. Nakamoto added a new money on top of the 
pyramid of monies, as little inflation as you can have, given that the system must pay for itself.

The story of money since say 1900 seems to feature mainstream thinkers, among them Georg Friedrich Knapp, 
John Maynard Keynes, and Milton Friedman, as well as supposedly marginal thinkers such as Sylvio Gesell, 
Abul A'la Maududi,\footnote{Maududi has revitalized interest free finance in the 1930-ies and by now
that approach seems to be effective, stable, and self-supporting.}
and now perhaps Satoshi Nakamoto. Gesell, and Nakamoto have doubts in common concerning the ability of 
governments and central banks to apply their powers of monetary 
governance in a way that advances the interest of all citizens. 
Gesell's views can be understood in terms of sometimes promoting (positive) inflation, Maududi's approach 
works best without inflation, and Nakamoto's approach points to a preference for
negative inflation, that is deflation.

So what is the mainstream story of money?  We are not economists and should
not even try to answer that question. The informational turn implicit in Nakamoto's conception of money, challenges 
every informaticologist to rethink the matter, in spite of the significant ``risk'' that the story of money 
has been written in essence already. In this paper we will try to 
formulate a conception of money that seems to be in some sense beyond that of Nakamoto.\footnote{%
Clearly we cannot be sure of that. But is seems that Nakamoto's objective has been to improve money, and we
suggest a kind of change, although we will stay very close to what we understand as being the essence of 
Nakamoto's proposal.}

\subsection{Survey of the paper}
The paper starts with an informal specification of what we will call a Nakamoto architecture. That specification
abstracts from many specific design decisions that have been taken in the original and 
subsequent Bitcoin client implementations. In order to have a readable exposition we made simplifications and 
by consequences some over-specifications as well. Bitcoin is an implementation of a specification that can be obtained from the 
Nakamoto architecture by appropriately adapting the technical details of transactions and mining.

Having introduced  Nakamoto architectures we turn to the concept of money.
A specification of features of monies provides a setting where combinations of these features
determine types of so-called near-monies.  Some near-monies may qualify 
as monies as well. Bitcoin is considered a near-money while its classification as a money is
plausible but may also be contested.

In Section~\ref{Wim2} the concept of an exclusively informational money (EXIM)  is proposed. We will not
consider Bitcoin to belong to the class of EXIMs,  although it plausible that some observers may 
consider Bitcoin a realization of the requirements that we have imposed on an EXIM. In order to classify Bitcoin
as a near-money we also introduce the notion of a technical informational money (TIM) and subsequently
Bitcoin is classified as a TIM. 

EXIMs are beyond Bitcoin (and other TIMs), and perhaps no single (near-)money that has currency to 
date can be considered an EXIM. EXIMs are more distant from conventional monies than
most commentators seem to consider Bitcoin to be. Under the name Bitguilder a hypothetical technical copy of Bitcoin
is introduced. Compared with Bitcoin, Bitguilder incorporates a different
perspective on its operation and on the service it offers. In contrast with Bitcoin, Bitguilder is considered an EXIM.

In Section~\ref{Security} an informal analysis is made of the security concepts that a user needs to have at hand in order to
be able to claim reliably that he or she can securely operate an EXIM. The level of security required for the use of 
an EXIM exceeds the security level required for making use of a TIM.

One may operate Bitcoin at the level of security required for an EXIM by means of a policy called 
{\em security by forward  physical separation}. This is a remarkable virtue of the Bitcoin protocol: it can be operated
securely with an amazing minimum of computer security precautions. It is far from obvious that conventional electronic
banking can be performed by a customer (user of the banking services) at the same security level.

Because  security by forward  physical separation is cumbersome and costly, and highly impractical for quick transactions,
attention is paid to the security implications of giving up the strategy of security by forward  
physical separation in favor of the application of automated 
processes on a single universal computing platform. 

In principle every TIM allows a transformed version that constitutes an EXIM. Bitguilder is the result of 
that abstraction when applied to Bitcoin. The
perception of rights and expectations for an EXIM user differs significantly from that perception for a TIM user. 
These differences are surveyed in the case of Bitguilder and subsequently a preliminary
survey is given of plausible extensions that will strengthen Bitguilder. 

%
In order to complete the picture on Bitcoin, two topics are further discussed: 
(i) using the notion of natural kinds as a tool for understanding how Bitcoin might
play an important role in the evolution of monies even if as a system it eventually fails, (ii) a dual system architecture
that combines an EXIM and a TIM which seems to solve major conceptual problems in the area of interest prohibition.

Finally some aspects concerning Bitcoin are discussed which are unrelated to its classification as a TIM: 
the experimental status of Bitcoin, paradoxical aspects of Bitcoin
and the open source project supporting it, and some aspects of the Bitcoin business case. 

In the first appendix a connection is discussed between promises (as conceived in recent theoretical work)
and actions in a peer-to-peer system that realizes the Bitcoin protocol. In the second appendix notational issues 
are dealt with that arise when using Bitguilder (or Bitcoin for that matter) as a unit of account. 

\subsection{Justification of the complexity of our approach}
For potential readers with a casual interest in Bitcoin this paper may appear outrageously complex and lengthy 
and overly decorated with terminology and notation. Our justification of these complexities lies in the following 
requirements that underly this work:
\begin{enumerate}
\item A wide variation in conceptions of money must be taken into account, thus reflecting that the concept of 
money is by no means trivial.
\item Whether or not Bitcoin is money (a question stated for instance in~\cite{Barok2011}) cannot be determined
with certainty  at this stage, both options need to be taken into account.\footnote{%
For the legal status of Bitcoin and its regulation see \cite{Kaplanov2012}.}
\item No assumptions about ``what Bitcoin is supposed to deliver'' or ``why certain design decision were made'' beyond
some statements by its originator in the 2008 white paper can be made. Moreover, the potential role or roles of
Bitcoin must preferably not be understood or explained in terms of a so-called libertarian view on society, 
or in terms of any revolutionary political or philosophical position, in spite of the fact that the yet unknown history
of Bitcoin's origination may be productively guessed in those terms.\footnote{%
This restriction implies that Bitcoin should be considered
from a wider range of societal views than would have been the case when its designers had delivered a complete ideological
story around it.
Bitcoin is far more interesting because of this open ended character. That alone makes the Nakamoto  anonymity move 
a very fruitful one, besides its major virtue of removing a single point of attack when 
Bitcoin comes under legal or political fire.}

\item Whether or not acquiring legal tender status is a step forward in the evolution of Bitcoin 
(or rather the co-evolution of Bitcoin and the rest of the world) cannot be decided, 
both options must be left open.\footnote{%
In~\cite{Abhishek2013} the case is made that India should work towards legal tender status for Bitcoin.}
\item No prejudice about the economic future of Bitcoin or about what use of it is most relevant must stand in the way of
an attempt to understand what role it might play.
\item Whether Bitcoin's best future (if any)  is to coexist as a money (or near-money) within a bundle of different 
(near-)monies or to serve all purposes of monies at the same time cannot be predicted at this stage. Both options
(apart from future extinction of Bitcoin) must be taken into account.
\item Assertions that might be understood as an invitation to illegal behavior must be avoided. (Stating 
that Bitcoin is an EXIM implies that capturing someone else's control over an address and the corresponding 
quantity of BTCs would be permissible. Now in the eyes of an opponent of that view, currently the majority view on
the matter so it seems, such a statement would violate this constraint). 
\item Such assertions cannot even be made as a thought experiment. (Rather than hypothetically classifying Bitcoin
as an EXIM, we will classify a hypothetical clone of Bitcoin as an EXIM. That thought experiment with a hypothetical clone 
cannot and should not be misunderstood as an invitation to consider Bitcoin an EXIM, not even temporarily).
\item No premature conclusions may be inferred (and used) from the ongoing and sometimes spectacular events on the 
Bitcoin exchange markets.\footnote{%
Comparing Bitcoins to shares in a company and understanding a sudden and sharp decline in Bitcoin/Euro rate as a sign of 
Bitcoin's impending demise is probably misguided in spite of the high 
plausibility that causality does work in the opposite direction. 
In connection with Bitcoin bubbles the following: 
the `Dutch' tulip bubble has not made tulips extinct, a stock market bubble 
leaves the stock market in place, a housing price 
bubble and its subsequent collapse creates difficulties outside the housing market which itself
always survives such an event. 

The end of a Bitcoin bubble may bring about a healthy shake-out of opportunistic
miners, and dramatic Bitcoin rate fluctuations may constitute important stress tests for Bitcoin client software. 
Repeated negative shocks in Bitcoin
value (as measured in Euros) may convince hoarders that Bitcoin holdings should better be used or be changed 
back to other monies which are either more stable or more useful for other reasons, thereby creating transaction volume,
a quantitative measure that undoubtedly counts for the survival of Bitcoin. DDoS attacks on Bitcoin intermediaries induce a 
mechanism of natural selection where those agents will survive whose defense against such attacks is most effective.

Whether or not Bitcoin is deflationary cannot be determined from its rate with monies that are known to be inflationary. 
Bitoin's rise in value may constitute a, possibly misguided, market consensus that conventional monies are more inflationary
than financial authorities are ready to admit. 
Bitcoin's future may be that it 
backs the Euro, or that it facilitates nano-payments, or Bitcoins may serve as collector's items forever. Survival 
as well as eclipse of Bitcoin may take place in unpredicted ways. Some market oriented blogs and columns are informative,
for instance~\cite{Liew2013} provides an informative and quantified vision of possible developments on the 
Bitcoin market from the perspective of a venture capitalist.}
\end{enumerate}

\section{Abstracting from  Bitcoin: the Nakamoto architecture}
\label{NakArch}
We will start with formulating an abstract view of Bitcoin, missing out on many details, 
but capable of being implemented in many ways differently
from the current realization known as Bitcoin. With the following listing of 
requirements we provide a specification of what we will
call a Nakamoto architecture for an informational money (say M).\footnote{%
The B-money of~\cite{Dai1998} contains many elements of what we are calling the Nakamoto architecture.}
 Bitcoin is assumed to be a realization of this architecture. The architecture
resulted from reverse engineering from information found in literature about Bitcoin,
rather than from our design effort. The idea is to capture Nakamoto's software design in an 
abstract way and in  such manner that significant modification of the 
Bitcoin design may still lead to a realization of the same architecture.\footnote{%
The style that we will use for this description is an informal version of the style used for patent 
descriptions: a numbered sequence of claims about an invention. 
The Nakamoto architecture may be considered a non-trivial (see also \cite{BergstraK2007}) 
software invention with Bitcoin as a realization, 
that might have been patented had Nakamoto wished to do so.

In \cite{Steadman2013} a survey of Bitcoin-like systems is given. We have not made an 
attempt to find a common abstraction for the systems covered in that paper. For the topic of P2P informational monies to 
become amenable for theoretical investigation it may be necessary to develop a more flexible approach to the 
Nakamoto architecture, so that it becomes generic and parametrized by mining features such as proof of 
work and proof of stake.}
\begin{enumerate}
\item Those who deal with M must participate in a peer-to-peer network, say N$_M$.
\item Participants of  N$_M$ split in two groups: users and miners (miners must be users as well). 
Users profit from the  N$_M$'s existence because it provides them with the functionalities that they need. Users interact with
the network though software clients, which may be but need not be open source programs.
\item Besides users there are (perhaps many more) indirect users, who make use of an intermediary agent, who is a (mining or 
non-mining) participant, and who acts on behalf of a collection of indirect users as an operator on N$_M$. Indirect users 
may be served by different intermediate agents. These intermediate agents are delivering participation services to indirect users.
\item Miners are participants who work for the system, mainly to ensure distributed database integrity.
\item Miners are rewarded by quantities of the unit $u_M$ of M. These rewards may either embody new money of
may be taken from users as fees for validating their (outgoing) transactions. Rewards are made available to the account from
which a miner operates.
\item All circulating quantities at any time have their origin as mining rewards at some stage.
\item There is no penalty system in the Nakamoto architecture, but that is a feature that can be easily added.\footnote{%
There could be a penalty on a proven double spending attack, namely that both transactions fail and the 
quantity that could have been transferred flows back to the system.}
\item Miners work in regular competitions to produce a transaction chain. 
Miners may group transactions together in blocks but that
is merely a matter of implementation.
\item User can have access to so-called accounts, which are natural numbers below given maximum 
 $k_M$ a parameter of M.  Access is obtained at the same time with an account by creating (a user action) 
 a public/secret key pair, say $(a,b)$ with a fixed cryptographic technique that is suitable for digital signatures. Then $a$ is an 
 address, which the creating user may publish to some other users or indirect users and $b$ is 
 the secret key through which the user has access to account $a$.
 \item Accounts give access to non-negative rational quantities of the unit of M. In practice such quantities may be multiples
 of a predetermined smallest fraction of the unit, but when needed the network can 
 be adapted so as to permit further division of  its unit.
 
Logically speaking at any instant of time a function $q$ is maintained in public that assigns to each address $a$ 
that has been generated an amount $q(a)$ accessible via that address for any agent who has access to the corresponding
secret key.
\item A pair $(a,q(a))$ that represents the result of the chain of validated transactions concerning  
an address $a$ at some moment is considered an informational coin in existence at that moment.
Informational coins have an age, found by determining the time elapsed since the last transaction involving its 
address.\footnote{%
This seems to depart from Nakamoto's use of the terminology. Alternatives to this definition 
are to consider transactions as coins or to include
secret keys in coins, or to have coins signed by secret keys.}
\item \label{TRscetch} Users have a single action at their disposal, effecting an outgoing transaction: 
to transfer a quantity $q$ $u_M$ from account $a$ to another account $c$.\footnote{%
Thus as a consequence of the previous item: $a$ is
a private key with a corresponding secret key $b$, only known to the user, who was responsible for 
creating the pair, and who is always able to create further key pairs, in a finite world of bit 
sequences that is practically inexhaustible. The
public key cryptography used in the system is the same for all users and is suitable for using $a$ as a digital signature.}

\item \label{TRdetails} A transfer effected by user $U$ works as follows: 
\begin{enumerate}
\item $U$ creates a random nonce (bit string of a fixed length; the length being determined 
by a parameter $n_M$ of N$_M$) $r$ which is supposed to be unique for the transaction during the entire life-cycle of N$_M$,
\item $U$ chooses a fee $f$ as a reward for the forthcoming validator of the transaction, 
(there may be prescribed fees, suggested fees, or minimum fees, validation may be refused when a fee is considered too low),
\item $U$ checks that the informational coin $(a,q(a))$ satisfies $q(a) \leq q+f$, (otherwise the transfer cannot be validated),
\item $U$ posts on the internet (intending to reach all of N$_M$) the message $\texttt{sign}_b(r,a,q,c,f)$ in such a way that all
participants can take notice of that posting. 
\item By inspecting the growth of the transaction chain (see below), 
$U$ can confirm that the transaction has been validated; an informal rule of thumb 
determines how many successive validations of other transactions 
(effected by arbitrary users) must be observed by $U$ for $U$ to be sure 
(sufficiently sure for practical purposes) that the transfer has been successful.
\item In theory (though with a probability decreasing exponentially in the number of later transactions that were 
subsequently confirmed) 
at any moment $U$ may find out that a transaction that has been successfully validated loses that status.
\end{enumerate}

\item \label{TRmore} Transactions can be more involved than indicated in~\ref{TRscetch} 
and~\ref{TRdetails} above by featuring several input accounts and 
output accounts at the same time. Transactions always include a nonnegative fee expressed in units or a fraction thereof which is
meant to be used as a reward for the miner, or miners if they operate in a pool, 
who succeeded in validating the transaction. Price competition among miners introduces a permanent downward pressure on fees.

\item \label{disclaimer-1} DISCLAIMER 1. The description of transactions 
in~\ref{TRscetch},~\ref{TRdetails}, and~\ref{TRmore} above is an 
over--specification and a simplification at the same time. What we intend to express is that ``some notion of 
transaction is used that makes the design/architecture work'', but that provides not enough information. 
What is specified here is not meant as a
specification that must be met in each N$_M$ but as a simple example of how transactions might look like, but permissive
of minor modifications of the design.\footnote{%
If one asks a mathematician to explain what a number is, the same phenomenon of over-specification 
is to be expected. As soon as a concrete number, say 37, is
put forward as an example, one is confronted with a specific instance of decimal notation, 
and is led to the question: what really is a number if we forget about such implementation details? Continued discussion with the
same mathematician will unveil recursive occurrence of the same phenomenon at differential levels of abstraction. It is very difficult to
convey an abstract idea without some over--specification that can be declared inessential in hindsight much more easily than being
prevented beforehand.}

\item Besides (ordinary) transactions there are mining steps, which are a special kind of transactions. Again simplifying, 
abstracting, and modifying the actual working of Bitcoin, a picture of mining steps is as follows.\footnote{%
Important aspects of mining can be found in~\cite{CourtoisGN2013}.}
	\begin{enumerate}
	\item A mining step is a tuple $(d, n, m, g, h, P, s)$ where:
		\begin{itemize}
		\item $d$ is the account from which the miner is working,
		\item $n$ is the number of previous ordinary transactions that the step covers,
		\item $m$ is the external difficulty of the problem $P$. The value of $m$ is supplied by the mining algorithm, 
		primarily depending on the transaction chain, and possibly depending on a variety of other parameters, 
		in such a way that the validity of $m$ can be checked at any time by all users.
		\item $g$ is the (total) fee collected in the mining step, that is the sum of the fees given by the 
		preceding $n$ (ordinary) transactions,
		\item $h$ is the size of the coin that is newly created in the mining step; a fixed algorithm 
		determines how $h$ decreases with increasing $n$,
		\item $P$ is a mathematical problem, which depends on the preceding transaction chain, and grows exponentially in
		$n$ and polynomially in $m$; it is assumed that all users and miners can uniquely and efficiently 
		generate $P$ from these data,\footnote{%
		In Bitcoin the problem $P$ consists of searching $s$ so that when $s$ is 
		appropriately included a bit sequence generated from the preceding elements of the transaction chain, $n$, and $m$,
		the secure hash algorithm SHA--256 produces a value (a natural number below $2^{256}$)
		that lies is below a certain value which dynamically depends on $n$ and $m$, (as well as perhaps 
		on the time needed on average 
		for successful mining in that phase of the system life-cycle). With the 
		current state of the art in cryptanalysis of SHA--256, miners have no other option than to generate many candidates for $s$,
		each time evaluating the hash function until sooner or later they will find a sequence that works. The expectation value of time
		needed for this can be reasonably simply computed and grows exponentially in $n$.}
		\item $s$ is a bit string which constitutes a solution of  the problem $P$; again it is assumed that all 
		agents (in particular all users) can efficiently check whether or not $s$ solves problem $P$,
		\item as a transaction the effect of the mining step is that the informational coin at 
		address $d$ has its value incremented by $g+h$, (the miner earns the fees plus the newly created amount).
		\end{itemize}
	\item A transaction chain consists of a sequence of transactions and mining steps, where the following conditions (which
	can be efficiently checked by all participants) are met:
		\begin{enumerate}
		\item The first element of the chain creates an initial coin at the disposal of the first
		miner,
		\item All digital signatures are valid; this can be checked by all participants without knowledge of any secret key information,
		\item the $n$ of each mining step equals the number of ordinary transactions the precede it in the sequence,
		\item $g$ equals  the sum of all fees contained in the last $n$ ordinary transactions, in the transaction chain,
		\item all transactions and mining steps from the beginning fit together in such a way that transactions only extract coins from
		pre-existing coins that are sufficiently large,
		\end{enumerate}
	\end{enumerate}
	
\item Transactions and mining steps are grouped together in so-called blocks. A block has a length $n \geq 0$, and it consists
of $n$ transactions followed by 
\begin{enumerate}
\item a sequence number $k$,
\item if $k > 0$ a message digest $h$ of another block (the message digest function used for this purpose is a parameter of the architecture; the hashed block is supposed to be the preceding block in the so-called blockchain which is mentioned below),
\item a sequence of $n$ transactions, $t_1, ..., t_n$,
\item a mining step $ (d, n, m, g, h, P, s)$,
\item and the combination of these items signed by the secret key of the miner. 
With $d$ the account used by the miner, let $(d,e)$ be the public/secret key pair that the miner has access to.  
\end{enumerate}
Summarizing a block has the following form.
$\texttt{sign}_e(k,h, (t_1, ..., t_n), (d, n, m, g, h, P, s))$.
The task of miners is to produce blocks and to send these around to all participants. The workload of that task is dominated by
the effort required to solve the problem $P$ in order to create a mining step as the final component of a (yet unsigned) block.

\item The history of the system up to some moment consists of a consecutive sequence of blocks, where all transfers fit. This is called the blockchain. All active participants maintain a current version of the blockchain. This requires both extension of the blockchain by
new blocks that have come available, and replacement of a tail of the blockchain if a block has been generated.

\item  Assuming that some part of the transaction chain has been created, and is known to all participants, 
by being stored in their local database, then the mechanism of transaction chain extension works as follows. 

New transactions (yet unconfirmed by any miner)
are being created by users and sent around. Participants store such candidate transactions in a local database in 
the so-called transaction pool. 
In addition miners will group suitable subsets of candidate transactions (that have not yet been included in the current blockchain) together into a sequence for which they can produce a mining step that completes it.

Once that is done, say by miner $X$, the block sequence number and the digest of the existing blockchain
are integrated into these data and signed with the secret key that $X$ is using. Then $X$ sends around the new block
just finished to all participants. 

Upon receiving the new block 
they will all check that it satisfies the mentioned requirements and if so, the participants subsequently 
extend their instance of the blockchain accordingly, by appending the block just received, summing that its sequence number
is the successor of the last sequence number of a block in their current blockchain.

\item \label{second-new-block} Now suppose that somewhat later miner $Y$ sends around a longer 
ordinary transaction sequence ending in a mining step (that is a longer block), then  all participants 
will drop the last part of the chain, (that is the block that $X$ had produced and sent around),
and will make use of this later entry instead by appending $Y$'s block. A block that survives this competition is called winning.
A voting mechanism among users for determining the winning block at any instant of time is a parameter of the architecture.

\item Miner $Y$ of item \ref{second-new-block} above may be seen as an attacker who attacks 
the block that had been produced by $X$.
A miner, say $Z$, may succeed in selecting a sequence of transactions that is so long that it attacks two or more blocks. 
An attempt to do this occurs if the sequence number claimed for the block is lower than the number of the last winning 
block that all participants have in store up to the moment of the attack on the blockchain consensus mounted by $Z$. 
Such an attack is successful if the new block chain is correct, and if the total difficulty (determined 
using a method to combine the problem difficulty of consecutive blocks which is a parameter of the architecture) of 
the single problem that the miner had to solve exceeds the
sum of the difficulties of the blocks that are being replaced, to a degree which is a parameter of the architecture.\footnote{In principle
the entire block chain can be replaced by a miner who solves a very difficult problem $P$. In practice, 
say in Bitcoin, that is extremely unlikely, but methods to freeze initial parts of the blockchain, 
thereby preventing such extremely disruptive attacks, may be needed on the long run.}

\item DISCLAIMER 2. Corresponding to the disclaimer mentioned in item \ref{disclaimer-1} above, it must be stated that
the above presentation of 
the mechanism of mining and the syntax of blocks and blockchains constitutes an abstraction and an over--specification at 
the same time. Variations on this theme are meant to be captured under the umbrella of the Nakamoto architecture.

\item Miners are constantly trying to group collections of new and not so new transactions together in 
order to append mining steps in the hope of creating a winning block that secures them control over combined fees plus
the newly coined reward for successful mining. 

\item \label{patience} Supposing that user $V$ is in control of account  $c$, then $V$ can check by means of $U$'s public key $b$ that
the transfer $\texttt{sign}_b(r,a,q,c,f)$ is authentic. Nevertheless $V$ must wait until at least one miner has completed 
so much of the validated transaction history of the system, including a log of the mentioned transaction, that $V$ may safely 
believe that the transaction took place and will not be reversed. 

This calls for a somewhat arbitrary threshold $k_c$ ($c$ for confirmation) 
where $V$ awaits validation of $k_c$ subsequent transactions before it performs any activity conditional upon having
received $q$ as a payment beforehand.

\item In spite of the soft guarantees implied by item \ref{patience} transactions that seem to have been 
performed are reversed if some miner  
succeeds in recomputing the history into a new blockchain 
that is more convincing to the majority of users. It will prove progressively harder for a miner to
succeed with this effort if a longer part of the blockchain must be rewritten.
\item A miner who is in control of over 50\% of the computing power of all 
miners combined has a chance of succeeding in rewriting the entire blockchain.
\item If no clear majority can be found when voting for blockchain extensions 
and two or more transaction histories are maintained simultaneously 
by different groups of users a so-called fork has arisen and some mechanism 
must be in place to detect this problem and to terminate that state of affairs. 
A fork is a fundamental problem, if it cannot be resolved it will disable the system forever.
\item Forks arise from so-called double-spending attacks which take place when a user transfers fractions
of the same informational coin to different accounts simultaneously or almost simultaneously. The idea of double-spending
is to receive a service in return from the agent in control of at least one of these accounts before they find out that the
transaction won't be validated after all. 

The rationale for the mining mechanism stems exclusively from the need to prevent successful double-spending 
attacks and to have the prevention of those attacks effected by participants of the peer-to-peer network 
without the introduction of a single point of failure.

\item Because miners will be rewarded with an informational coin, and because validation itself is a trivial task, 
the need arises for miners  to be in a significant competition for that money
in such a way that successful completion of the competition can be checked by every user. The work to be performed for
the competition is called the mining contest assignment. 
\item The interaction between participants is constrained by a protocol, which essentially consists of the following elements:
\begin{itemize}
\item Data formats for signed transaction (sent around) and unsigned transactions (appearing as entries in the transaction chain), 
\item A common standard for digital signatures, a type of problem that (a) miners can both generate and 
competitively solve, and (b) for which users can check the correctness of the problem solution.\footnote{%
At the abstraction level of a Nakamoto architecture we will not specify the type of problems and their solutions. Clearly in the 
preparation of Bitcoins a working realization of these constraints has been provided.}
\item A data format for solutions to the competition that miners perform.
\item A voting criterion to be used by all users indicating which miner has been victorious.
\item A yield expression  that determines how much reward for the winner of a mining competition is generated. 
\end{itemize}
\item \label{last-item} The earnings of miners in terms of new money diminish in time in such a way that an asymptotic growth of the
total quantity of units circulating in $M$ is built in in the system. The limit value thus obtained serves as a maximum $C_M$, 
or better upper bound. $C_N$ is a key system parameter and its value is encoded in the yield expression in the protocol. 

This maximum is reached along a predictable curve and within a predictable expected time, say $T_M$. 
Both the shape of that curve and the last
moment of new unit creation is determined by the yield expression.

\end{enumerate}

\subsection{Preliminary specification, open for improvement}
Bitcoin is the open source developed offspring of Nakamoto's own implementation of what we have 
just described as the Nakamoto architecture specification. Bitcoin has been live since early 2009, when Nakamoto offered
a first software client ready for public downloading under an open source license.  The unit of Bitcoin is denoted BTC.

We assume that Bitcoin, as it is in existence to date, complies with the architecture just specified. It remains to be 
seen whether or not the~\ref{last-item} ``axioms'' mentioned above provide a meaningful abstraction of Bitcoin.
In this specification of a Nakamoto architecture several aspects have been intentionally left open for further refinement,
as design decisions to be taken when developing implementations. 

For Bitcoin $T_M$ is about 2040, the smallest fraction supported by current clients is $10^{-8}$ BTC, 
$C_M$ is about $2,1 \cdot 10^{7}$, the curve is sloped downward in such a way that early 2013 about
half of the BTCs that are expected to exist have already been mined.

This specification is at best  a preliminary specification in need of review and improvement. For obtaining an
 implementation of a Nakamoto architecture several additional choices must be made. We mention some:
\begin{itemize}
\item The problem family from which miners take their problems, and the checking mechanism applied by users. 
\item The grouping mechanism used by miners to collect numbers of transfers ready for validation.
\item The yield expression, voting criterion, and the circulation maximum.
\item Additional constraints imposed on mining users (a need for certification cannot be excluded; 
additional measures against some miners becoming too strong may be necessary),
\item Methods for freezing an initial number of blocks in the blockchain so that these blocks a made immune against
any later attacks by computationally excellent miners.
\end{itemize}

The circulation maximum constitutes the main defense of the Nakamoto architecture against inflation. 
For each quantity it is known in advance which fraction of the entire circulation it will eventually represent. 
Only that fact produces a reason for a user or an indirect 
user to trust that it has some value, as measured from outside the system, to have exclusive 
access to a quantity of units via an account, by being uniquely in control of the corresponding secret key.

\subsection{Is Bitcoin a hype?}
At the time of writing these lines a BTC (Bitcoin's unit) is rated about 190 Euro\footnote{%
Often BTCs are sold and bought by agents running a participation service in behalf  of indirect users.
Indirect users may be (direct) users at the same time and may have BTCs that they own indirectly 
via a participation service transferred
to an account that they are controlling. In this way users may get access to informational coins 
without having been successful in mining.
That is important because mining has become extremely competitive and in need of huge investments in a few years time only.}
 and many commentators speak of a hype with a high risk of collapse driven
by opportunistic buyers who gamble without having an economic rationale for their actions.\footnote{%
When revising these lines two days later the rate is down to 70 Euro. At the time of preparing version 3 of the paper (end of December 2013) A BTC trades for 500 Euro, after having been at a top of about 900 Euro in November 2013. These high
rates constitute an incentive for issuing warnings to the general public against Bitcoin for financial authorities throughout the world.}
We will make an attempt to estimate the expected value of a BTC.
We compare Bitcoin with a high tech startup which will either become dominant on its 
market or it will fail.\footnote{%
In~\cite{Yermack2013} the case is made on the basis of market data that Bitcoin behaves like an 
Internet Stock from the late 1990s rather than like a bonafide currency.}
We assume that Bitcoin cannot be bought by a competing money. 
A probability needs to be found for success. Given the technical novelty of the system, and given its
minimal level of institutionalization we choose a seemingly low probably for its success: one in a 100.000. Now if Bitcoin 
survives till 2040 and is very successful then it may replace old-fashioned technologies and it may, say, 
represent half of all money world wide. That is a difficult estimate and we will use $10^{14}$ Euro as a guess. 
That produces an estimated BTC value in of about $10^{14}/2 \cdot 10^{7} = 5 \cdot 10^{6}$ Euro with a 
probability of $10^{-5}$, that is 50 Euro. 

The risk of loss caused by capture in an attack, or simply caused by incompetent
bookkeeping of secret keys can be ignored as it is compensated for by either increased assets of
other users, including first of all the attacker, in case of capture, or by an increase in the value of all 
remaining BTCs in case of loss.\footnote{%
Obviously, if a user has no effective strategy for keeping secret keys secret, buying a BTC and transferring that to
an account in control of that user is not advisable.}

Of course this estimate can be adapted progressively as more information becomes available, and 
both the probability of survival may increase, perhaps until a sudden collapse, and the total value estimate 
that will be reached may increase, the latter being independent of Bitcoin.

\section{What is money I:  functional dimension}
\label{Wim}
The question ``what is money" seems not to have a definitive answer.\footnote{%
In~\cite{Bergstra2013a} an attempt was made to provide a definition of money. To that end a dedicated theory
of definition is developed. Unfortunately getting rewards, in terms of clarification of the essence(s) of money, 
compensating for the additional weight of a detailed theory of definition is not straightforward, and may not have
been achieved in~\cite{Bergstra2013a}. 
Nevertheless we believe that having a theory of definitions at hand
when defining money is equally necessary as having a theory of real numbers one's disposal when 
defining differential equations.} 
The question can be rephrased, made more inclusive, and for that reason simplified as
follows: ``what is a money?"
Here ``a money'' combines in a single concept the system (institution, network, or organization) that permits
certain quantities to circulate through the hands of a community of users for a variety of purposes
and the individual occurrences of these quantities as these are flowing through the network.

We will take for granted that the physical carrier of a money is shaped by way of information technology,
and that valuable physical coins and nicely printed banknotes have already ceased to exist, or play
a marginal role only. Those
features do not distinguish classical monies from Bitcoin like monies. What makes the concept of money hard to 
define is that commonly a certain combination of characteristic features is used as a criterion (if those are
met that justifies speaking of money, at least of an initial concept of money) 
while at second inspection more is needed because one invariably needs some refinement of 
the initial conception of money when dealing with its realizations.

Another perhaps more provocative way to express the difficulty of defining the concept of money is to say that
each particular money in some sense gives expression to a bundle of political and economic viewpoints. There are
many such bundles and  for that reason there are many different monies. The proponents of each particular 
money have great interest
in portraying that particular form of money as being natural, that is as being obviously 
one of the forms that all money can take.

\subsection{Monies and near-monies}
A formidable complication for the student of monies (including new and exotic monies) 
stems from the fact that what may count as a money for one observer need not qualify as a money
for another observer. We need an approach that takes care
of this complication in a principled manner. We will use the term near-money for a system of circulating items
that satisfies so many of the requirements (characteristics) of a money that it is a candidate for being money.
A money is always a near-money as well.

Then we assume agreement on what near-monies are, whereas the
question which near-monies are proper monies may split observers. We assume in particular that Bitcoin 
is a near-money. A systematic analysis of Bitcoin is required before the question whether or not 
Bitcoin is a money can answered in a satisfactory and informative way. Typically, what some consider the 
greatest strength of Bitcoin, that it has a built in defense against inflation caused by increased circulation, 
still leaves room for serious doubts in the minds
of other observers regarding the claim that Bitcoin can serve as  a store of value.

We will assume that the term near-money is ideologically, legally, and politically, sufficiently unprotected
to be applied without hesitation to Bitcoin, an assumption that we cannot make about the more restricted 
concept of money itself. This classification issue  for novel near-monies like
Bitcoin matters in particular in financial architectures 
where a bundle of different near-monies coexist, each having its own purpose and function,
and where some specific philosophy of money is applied to determine which of the 
near-monies count as monies.\footnote{%
The same monetary architecture (bundle of near-monies) may be analyzed with different 
philosophies of money in mind. This degree of freedom in analysis 
may be compared with the field of software engineering where different architectural methodologies
such as object orientation, aspect orientation, and agent orientation, can be approached from the perspective of different 
programming paradigms such as imperative programming, functional programming, and logic programming.}

Below we will use barter catalyst as an equivalent for near-money because all near-monies seem to be means of exchange
that facilitate trade. Barter catalyst expresses that clearly.

\subsubsection{Classical characteristics of circulation and storage}
Classical characteristic features which can be found in any book on money\footnote{%
All characteristics/features/requirements are presented as pertaining to 
monies with the understanding that a completely rigorous
exposition would first speak in terms of candidate near-monies, then spell out thresholds in terms of compliance with
the characteristics/features/requirements that lead to a classification as a near-money, and finally provide a 
description of one or more philosophies of money each giving rise to a specific qualification of some near-monies as monies.
In Paragraph~\ref{IntProh} below we will outline how a philosophy of money based on interest prohibition might conceivably 
opt for a dual-near money system with a Euro-like near-money and a Bitcoin like near-money harmoniously coexisting, and
not classify the Euro as money but merely as a near-money while the Bitcoin-like near-money is viewed as a money.}%
in some form or another are these, listed in arbitrary order:\footnote{%
Economists differ in the relative importance they assign to these features. `Pessimistic' commentators of Bitcoin
often base their opinion on the implicit assumption that being a means of exchange is 
the primary functionality of a money. We will try not to rely on such debatable preferences.}
\begin{description}
\item[\em Barter catalyst.] (Usually referred to as means of exchange.\footnote{%
We prefer the phrase ``barter catalyst'' to the conventional phrase ``means of exchange'' because the latter
suggests that exchange needs such means which is certainly not the idea with mechanisms like Bitcoin. Bitcoin may
support (catalyze) trade without monopolizing it.}
The circulation of  money-items\footnote{%
In connection with Bitcoin some authors use the convention that Bitcoin refers tot the system while Bitcoin refers
to the items in circulation. We will not apply that convention.}
within a community of its users
 is helpful with enabling barter between members of that community, where (i) barter extends to access
 rights to  other classes of goods and services than those immediately connected to money and (ii) the 
 barter mechanism has some (limited) flexibility to 
 extend to (access to) new goods and services. 
 Barter catalysts relieve participants from the constraints of the so-called coincidence of wants.)
 \item[\em Unit of account.] (The ``money'' induces a notational scheme for indicating (accounting) values 
 (costs, prices, expected costs, contemplated prices, valuations of assets, productivity of community 
 or of a specific  part of community)
 in the absence of the underlying money items. A proper unit of account is also helpful for barter 
 participants but
 it does not help against the limitations resulting from the needed coincidence of wants. A practical unit of account
 is helpful for establishing a coincidence of wants by making different offerings more comparable.) 
 \item[\em Store of value.] (Non-circulating money-items can circulate in the future and that potential for future circulation
 represents wealth or value that an individual participant can use in
 the future. This feature creates the accumulation of (stored) money as an opportunity as well as 
 an independent ambition of a community member.)
 \end{description}
 
\subsubsection{Characteristics connected with scope, purpose, and status}
Scope, purpose, and status are non-circulation based  aspects of monies
may also understood as further requirements to be met by a money. There is no assumption, however, that all of
these requirements can be met simultaneously by the same money design. Here is a survey of such additional aspects:

\begin{description}
\item[\em Legal tender status.] If a money is acceptable for a state for paying its taxes the money has legal tender status
in that state.
\item[\em Rigid quantity management at issuers discretion.]  The quantity of money at any instant of time is given by
the design of the money in advance and for that reason to some extent predictable.
\item[\em Flexible quantity management at issuers discretion.]  The quantity of money is managed by  
the body in charge of issuing the money, in a flexible way in order to promote the functionality of the money for its
entire user community.
\item[\em Geographic area universality.] A money is universal in a geographic area if it can be used for all 
exchanges and investments in that area (of course with an exception of transactions carried out on purpose with other monies).
\item[\em Exclusive geographic area universality.] A money has exclusive geographic area universality if it is the only
universal money in that area.
\item[\em Degree of freedom.] A money must allow its participants some freedom of use. 
(The way in which money will circulate 
through its user  community cannot be predicted in advance. In fact the unpredictability of that circulation adds
to the functionality of the medium.)
\item[\em Functional communication protocol.] A money may be dominant in some area or sphere of life. 
Where that is the case it plays a role in the
organization of a community and it can and should be judged according to the way that role is played. (From
some perspective a money is merely a communication protocol between community members that helps them
achieving other and more important purposes than the accumulation of monetary value such as the 
completion of important public works).
\item[\em Enabler of human rights provision.] To the extent that a society must allow its members access to certain
goods and services, it may organize a market where such goods and services can be traded for ``the money
under consideration''  under the condition that all members of the community will be provided with enough money
to serve their minimal needs, and will receive support for their spending strategy if the cannot do without.
\item[\em Instrument of internal governance.] A money may be dependent on a variety of parameters that can be modified
in principle without significantly changing the design of the money.\footnote{%
In Bitcoin such parameters are: (i) the rate at which the return on successful block mining fluctuates
(it might be made negative if the Bitcoin quantity is considered too voluminous and start increasing thereafter etc.); 
(ii) the rate at which block mining becomes more difficult with increasing block difficulty; (iii) the specific cryptographic
technology (SHA-256, see~\cite{FIPS2008}) for generating and checking proof of work); (iv) elliptic curve cryptography 
(ECDSA see~\cite{JohnsonMV2001}) for public key cryptography
based authentication (by means of digital signatures) of transactions; conventions for 
freezing initial segments of the block chain (for creating 
certainty that transactions cannot be undone in the future, and to simplify the verification task of participants.}
Adapting these parameters with the intention to improve the money's functionality (performance) on other criteria
is a matter of governance. Such adaptations are clued interventions. 
A money may be assessed according to the scope and effectiveness of options for interventions
that it offers its governors.

\item[\em Instrument of external governance.] Different currency units (large groups
making use of different monies) may allow interventions that in turn may compensate for discrepancies between
the performance of different currency units.
\end{description}

\subsubsection{Information technology enabled characteristics}
The following characteristics of a money may be present as features following from the technology that has been used
for its realization.
\begin{description}
\item[\em No single point of failure.] A money exhibits no single point of failure if each participant, irrespective of its
role can stop participating without causing more damage than a moderate slowdown of performance.
\item[\em Enabling non-reversible transactions.] A money enables non-reversible transactions if money can be transferred
in such a way that this cannot be undone otherwise than autonomously by the receiver.
\item[\em Enabling anonymous transactions.] A money enables autonomous transactions if transactions can be performed
in such a way that no link between sender, receiver, amount, and time can be found by an external 
observer though inspection of data that reside in the system.
\item[\em Autonomous transactions.] A money provides autonomous transactions if participants cannot be 
forced to perform transactions and cannot be prevented from performing transactions.
\end{description}

\subsubsection{Requirements on money concept formation}
Here we will outline several additional constraints that might be imposed on a concept of money, or more precisely,
on the context in which concepts of money are developed. Currently used
world-monies such as the Euro manifestly don't satisfy these conditions. These requirements may or may not  be met
by Bitcoin depending on one's viewpoints of Bitcoin. However, the four requirements listed below  
are met by its hypothetical clone Bitguilder which will be discussed extensively below.

These constraints must be understood from the perspective that different monies may coexist, and that compliance with
specific requirements may be appropriate for some monies and inappropriate for other monies.
\begin{description}
\item[\em Full definitional independence.] Money is so important that its definitions, 
and the practical implementation of such definitions, 
should not be made dependent on the intentions or wishes of any institutional agency, state, or government.

\item[\em Black--white self-governance.] Black--white self governance indicates that no-one else than users of a money will be
authoritative about the status of amounts. In particular no external party can distinguish between black money and white money.

\item[\em Legal tender insensitive.] Legal tender status ceases to be the holy grail of money design in the 
context of informational monies:
if the state (or any institution able to impose its views about money on a large group of citizens) 
wishes only to award legal tender status to a money the design of which compromises full definitional independence, 
black--white self--governance, or informational autonomy, then the designers of a money 
(and as a consequence of that its user community) must
have the freedom not to ask for or strive towards legal tender status.
\end{description}

\subsubsection{Missing aspects}
Rather than to adapt the above description of monies consecutive rounds, we will collect missing aspects, that
came to our attention only later, in revisions of this paper, in this section only to restructure the above section on monies
once a more complete picture has been found. We will now mention two aspects constituting a 
clear omission in the above listings.

\begin{description}
\item[\em Money as memory.] In~\cite{LutherO2013} the case is made that Bitcoin provides a viable match with those
economic views on money where memory is considered as a mechanism that participates in the market equilibrium.
In other words, where memory is very strong, money may be weaker. 
That money circulation and memory are glued together in Bitcoin is undeniable, 
the degree to which the information stored in the evolving blockchain suffices to meet 
economic needs may be open for discussion. Bitcoin may have the flexibility to allow 
customized compartments of limited circulation where memory functions are vastly 
expanded in order to accommodate specific requirements of communities of
economic agents.
\item[\em Money as defined by political, financial, and monetary authorities.] A range of national and international authorities 
produces regulations and statements about ``what is money", (and about the moneyness of Bitcoin). This leads to an extensive
secondary literature of which we mention~\cite{KrohnGS2013} only. It seems that no final conclusions can be drawn along this line of analysis as until now different bodies come to different view points about the same issues. On the long run we expect that the scientific judgement about the moneyness of any informational near-money and the majority judgement of regulating bodies about that same question will coincide.
\end{description}

\subsection{Assessing the Nakamoto architecture}
A money that implements the Nakamoto architecture can be assessed against the criteria on moneys mentioned above.
This assessment leaves options open while actual implementations of the architecture allow a more thorough 
assessment against these criteria. We assume that the architecture is successfully implemented by a money $M$, in
that case $M$ runs a risk of being weak as a barter catalyst and as a unit of account both due to deflation, it may be very
good as a store of value; $M$ may or may not have legal tender status, it admits rigid quantity management 
but no flexible quantity management, it will not have geographic area universality, and not exclusive geographic area universality, 
$M$ expresses degrees of freedom for its users, it provides a communication protocol, it is not 
an enabler of the provision of human rights, not an instrument of internal or external governance. $M$ shows no single
point of failure, enables non-reversible transactions. $M$ suitability for anonymous transactions depends on a combination 
of sophisticated usage and a ``friendly'' legal framework supportive of such forms of anonymity. In andy case this aspect must
be assessed with great care given an implementation at hand. $M$'s transactions are autonomous in as much as the
legal context is permissive of that form of autonomy. $M$'s implementations emerge in a setting which must be permissive
of limited definition independence, and of legal tender insensitivity. Experimenting with black--white self governance need not be permitted for designers of $M$, but having that option is an advantage.

\subsection{Competing monies and cooperating monies}
A specific money will feature some of the characteristics listed above but certainly not all. Different types of money differ in the
package of characteristics that are offered. The same package can still have vastly different implementations.

Very common is the coexistence of a bundle of monies each of which has exclusive geographical universality in a different
area. Just like nation states cover earth such monies do the same. Coexisting monies with exclusive geographic area universality
are cooperating and competing at the same time. Merging such monies to a single money is a complicated task because
cooperation and competition are different but equally important functions that may run out of phase, even when initialized 
harmoniously.

Bitcoin represents another package than known from conventional monies: 
no geographic bounds, no  claim to geographical universality in any area, let alone a claim to
exclusive universality. The store of value function in Bitcoin is paramount and is not compromised by 
quantity management.\footnote{%
In most conventional monies quantity management consists of regulated but frequent
quantity increases mostly with an inflationary side-effect. A reason for this tendency might be that 
deflation is often considered quite detrimental for an economy.} 

\subsubsection{Future evolution: rotation of the bundle of (near)monies?}
The bundle of existing monies may be moving away from horizontally juxtaposed monies
of the same type each featuring geographical universality in geographically disjoint areas. Newly created informational 
monies (e.g. Bitcoin and DigiCash to mention some), 
are increasingly geography insensitive and are combine by way of vertical juxtaposition.
Different geography insensitive monies may coexist in the same area (that is world-wide, or at least in all jurisdictions
permissive of the money) each serving different purposes, or simply competing when serving the same purposes.

Perhaps the architecture of monies will evolve through a 90 degree rotation, from a horizontal juxtaposition to a
vertical juxtaposition, with geographical universality becoming
uncommon into a bundle of monies where coexisting monies being of significantly different type.

\subsection{Technically informational monies and near-monies}
\label{TINMdef}
In the philosophy of money it is hard to escape from historic accounts that explain what 
monies have come to exist
and why it is today the way it has become. Such accounts may be unconfirmed and 
intelligible at the same time and
then constitute fragments of so-called conjectural history of money. 

A plausible (conjectural) path towards informational money leads along the 
automation via digital computers of known financial processes. This path leads through stages with hybrid monies
that are informational and non-informational at the same time.
Money is technically informational (TIM) if it may exist by way of stored information only. 
We will classify Bitcoin as a TIM.
In particular, abbreviating Nakamoto architecture to NA,  Bitcoin is an NA--compliant TIM.

The terminology that we have introduced induces other terms and phrases that seem less useful at first sight
but can still be imagined. In the case of a TIM it is plausible to start from a near-money rather than from a money. 
That leads to TINM for
a technically informational near-money. In Section~\ref{IntProh} we will discuss a setting where an EXIM and a TINM coexist,
and where it is both essential in that context and quite unconventional at the same time not
to consider the mentioned near-money a money.

\subsubsection{Managed TIMs and managed TINMs}
 If Bitcoin (or any TIM)
were modified  that some board of governors takes decisions about the size of the 
monetary base, which can be manipulated by making mining more or less productive\footnote{%
A negative premium on mining (a TIM's version of Tobin tax), 
 might still be compensated by transaction fees and would not bring the system to a halt 
 while effectuating a reduction of the BTC base.}   
 in such a way that an appropriate balance 
 between deflation and inflation is maintained in everyone's best interests 
 (as understood by the same governors), then that modified system will be a classified as a 
 managed TIM for which we will use the acronym MTIM.

A managed TINM is an MTINM.\footnote{%
In spite of this cumbersome naming MTINMs are very practical tools. Without management of some kind very 
unpredictable behavior may occur and missing some features deemed necessary for being a money may be helpful
in certain circumstances.}
We will not attempt to survey the forms of external management which can be applied to a TIM. Quantity management
is the most obvious approach. Imposing restrictions on transactions and holdings are  management options as well.

We will insist that an MTIM is also a TIM and that an MTINM is also an TINM. For a given TIM may different MTIMs
can be imagined that share the same technical kernel. That holds for an TINM as well.

\section{What is money II: autonomy and privacy dimension}
\label{Wim2}
The range from near-money to money may be understood a dimension is the space of possible monies. This is
the dimension of functionality. Monies are those near-monies that feature enough functions. The direction of this dimension
is that of offering more functionality to its users.

In this work another dimension of the space of existing and conceivable monies will prove to be
of equal importance: the degree of autonomy granted to a user in control of a quantity of (near-)money, and the degree of
privacy granted to a user's activities and capabilities (in particular access capabilities). 
 
Many near-monies can be understood as a circulation technology which allows being deployed
with different degrees of autonomy for its users. For a near-money that satisfies the 
Nakamoto architecture many different levels of autonomy of use can be imagined. 
We will introduce exclusively informational (near-)monies (EXIMs)
as those (near-)monies that permit a highest degree of user autonomy.

There is a difficulty in the terminology which needs clarification. It is possible that some legislation, or more simply a
personal view limits the degree of autonomy that is considered acceptable for a money. Thus, from certain perspectives, 
granting the users of a money too much autonomy may be inconsistent with the concept of money prescribed by that
perspective. Indeed, we will observe no contradiction if an exclusively informational money fails being a 
money in the eyes of an individual, of a  community, or of an authority because they consider it to be granting its 
users too much autonomy. We will not consider an exclusively informational money that fails to qualify 
as a money to constitute near-money for the reason of that failure alone. 
Near-moneyness only expresses some lack of functionality.\footnote{%
When needed we propose to call an exclusively informational money that provides its users too much autonomy, and that
is not considered a money for that reason a non-standard money. That naming should not be understood as 
having some negative connotation, however.}

\subsection{Rules of engagement: the AP sheaf around a TIM or TIMN}
We will focus on TIMs and TINMs (including MTIMs and MTINMs as subclasses). Given a TIM the technicalities
of circulation as well as the options for user action are fixed. Having said that, a significant degree of freedom is still given
by the variation of rights, obligations, expectations, norms, and conventions that can be imposed on the users of
that particular TIM. We will refer to a package of those aspects as the rules of engagement for the TIM or TINM.

Below we will provide a listing of such aspects. It appears that many selections from these aspects may be combined
in to packages of informally stated properties that go with a particular instantiation of a given TIM or TINM. 
In other words, given the system as a piece of technology, different combinations of rules of engagements can be imagined
as a framework in which the system can be used.

The autonomy/privacy sheaf (AP sheaf) of a TIM or a TINM collects all monies that can be obtained from that 
TIM or TINM by imposing on
its users a collection of obligations and granting them a collection of rights, as well as by assuming a set of 
convents for proper use. This allocation of rights, obligations, 
and conventions extents to outsiders to the TIM or TINM as well.

We understand certain rights, obligations, or conventions as non-informational properties of a money. For instance,
the obligation to tell the tax office what amount a user controls has no bearing on the informational content of the
money. Similarly the right to have a thief prosecuted adds some value external to the informational content of 
an informational coin.

Rather than to map out the full AP sheaf of a TIM or TINM, we will focus on an extreme element of that
sheaf which maximizes user autonomy. Such extreme elements qualify as exclusively informational monies.

We have not included anonymity in AP dimension because anonymity is a property that cannot be declared, it can be 
expected or hoped for, but that a system respects certain forms of anonymity 
concerning its ussers  a technical property of it that requires a proof.
Anonymity issues are equally pressing for TIMs and TINMs as for EXIMs and EXINMs.

\subsection{Definition of an EXI(N)M}
An EXI(N)M provides a distributed system that has the following 
functionality, while being compliant with the technical 
requirements  and with the essential conventions as listed below:\footnote{%
At this stage we are not in the position to classify the Bitcoin system as an EXIM or an EXINM, if only for legal reasons.

Below we will experiment with a hypothetical clone of Bitcoin, (pretentiously called Bitguilder
in order not to forget the Dutch Guilder that got lost when the Euro came into power). We would have preferred 
to use the name Bitcash instead of Bitguilder but unsurpisingly Bitcash turns out to have several 
established meanings already. 

As a thought experiment Bitguilder may be considered an EXIM. The consistency of that thought experiment,
which is implicitly claimed by its being performed, 
seems to imply but does not actually imply that Bitcoin itself can be considered an EXIM. 
Indeed, asserting that Bitcoin
as it currently exists, is an EXIM, implies the assertion that Bitcoins 
cannot be stolen (because they cannot be owned), 
an assertion which might be considered false by many and even illegal by some.}

The question whether an EXI(N)M is an EXIM or an EXINM depends on one's philosophy of money, 
just like the question whether a particular near-money is a money. Following the survey of features of money of
Section~\ref{Wim} the following features are needed from the perspective of circulation.

\begin{enumerate}
\item allows its participants to store and circulate informational value so that,
\item quantities of informational value may be exchanged between 
participants as barter catalysts, and 
\item quantities of informational value may be stored by participants 
so as to
preserve value available for use by the same participants at a later time of their choice. 
\end {enumerate}
Further the following technical requirements must be satisfied:
\begin{enumerate}
\item Quantities of informational value, so-called informational coins or INCOs,\footnote{%
We borrow from Bitcoin the term coin for any quantity that is immediately accessible in a homogeneous way (that is
via a single access or transaction) even if we feel the term coin is counterintuitive for such quantities.}  
are measured as non-negative rational numbers
(signed cancellation meadows~\cite{BBP2013} provide a useful syntax for these numbers) have no meaning 
except their interpretation in the meadow of rational numbers. It follows that INCOs can be denoted
with expressions of the form $q$ U with U standing for the unit of the INCO.
\item INCOs connect quantities with identities of some form 
(e.g. the public keys/addresses of Bitcoin).
\item Participants can only have access to INCOs depending on their having access to
additional informational items (INITs), for instance by having access to a secret key from a private/secret key pair.
\end{enumerate}
In addition to the technical conditions there are these rules of engagement. This package of rules determines an extreme
``point'' in the autonomy sheaf around a TIM or a TINM. Several subsets of this package may be as plausible or even more 
plausible given current conceptions of money:
\begin{description}
\item[\em Priority of access.] Participants can only have access to INCOs. There is no distinction between 
legitimate access and non-legitimate access.
Whoever can access an INCO is by definition entitled to its use for any purpose (which is permitted
as a form of use of any other INCO). 
Access cannot be disputable, it is a fact of matter or it isn't.\footnote{This does not imply that in a Bitcoin-like
EXIM double-spending attempts are permitted. There may be (but need not be) 
a penalty on such behavior 
because it requires other peers to do superfluous work in transaction checking, and the fact that at least one of
both fees is fake is problematic.}
\item[\em Access produces capabilities instead of rights.] Having access to an INCO is best seen as a capability. 
There is no right to access and access does not create rights.\footnote{%
A simple way to highlight the difference between access and ownership is found by means of a comparison
with classical currency in the 
form of coins and banknotes after an amount $a$ has been stolen from $P$ by $Q$, the access
to $a$ is with $Q$ but ownership of $a$ is still with $P$. Circulation of black money provides an example of a world
where access is prior to ownership, (at least as long as one does not
redefine ownership of $a$ through the codes of a criminal organization's
claim that the proper thief is now the real owner). Laundering is the opposite of stealing for money. Laundering $a$
after it has been stolen by $Q$ brings it into a state where $Q$ has ownership in addition to access.}
\item[\em Free access and transfer.] No participant can ever be forced or forcefully prevented 
to access (or transfer) any 
INCO which (s)he is capable of accessing.
\item[\em Autonomy of transfer.] Each transfer of an INCO by a participant who is 
capable of accessing that INCO must be a voluntary and autonomous act by that participant. 

In particular a (conditional) 
promise made by a participant  $P$ to transfer an INCO at a future moment (to another 
participant, say $Q$) cannot lead to an obligation to do so which goes 
beyond the  expectation that $P$ will act accordingly. As soon as some mechanism is put in 
place that enables $Q$ (or any agents except $P$)  to force a transfer upon $P$, for whatever
reasons, the entire system loses its EXI(N)M status.\footnote{%
This requirement matches the conventions on promises proposed by Mark 
Burgess in~\cite{Burgess2005,Burgess2007}.}
\item[\em Permissive for anonymous outgoing transfer.] If a user transfers an amount to another account,
the transferring  user will not be required to
disclose to any third party either the identity of the prospective receiver or the reasons for carrying
out the transfer. 
\item[\em Permissive for anonymous ingoing transfer.] If user acquires control of an amount, the user is not
required to explain to any third party what and how that has occurred.
\item[\em Abstraction from physical carriers.] INCOs cannot be identified with their 
physical carriers. Ownership of a physical carrier of an INCO does not imply 
ownership of the encoded INCOs (neither does it guarantee access).
\item[\em No ownership.] Access to an INCO dominates any notion of ownership.\footnote{%
The semantics of terms is problematic. 
In~\cite{Sweazy2012} a notion of ownership is used which is very similar to what we call access.}
\item[\em No theft.] As a consequence of the above: INCOs cannot be stolen, 
but in some cases they can be lost (if additional information
gets lost).
\item[\em No lending.] In addition transfer of access to an INCO transfers all possible 
rights and borrowing INCO's from another participant for temporary usage is impossible by definition.
\item[\em Control: the capability to access.] It will be said that at some moment in time a
participant controls an INCO or a quantity of $q$ U (where U 
represents the unit of the INCO), if the participant can access that quantity of INCO at that moment in time. 
The notion of control replaces the familiar notion of ownership. 
\item[\em Shared control.] Different participants may control the same
INCO at the same moment in time. In that case control is shared, otherwise it is non-shared or exclusive.
Sharing control with another participant may not be intended but it is any participant's responsibility to
prevent shared control if that is preferred.

In the case of shared access, the collection of participants having access to the said quantity of INCOs 
is called the controlling group.
\item[\em Exclusive control.] Control which is not shared is exclusive. In this case the controlling agent
(participant, principal) is the only one having access.

Although exclusive control is the 
clearest form of control and may give rise to the best predictable behavior of INCOs 
controlled by a participant, storing important value in INCOs under exclusive 
control may leave a participant quite helpless in case of significant problems.
\item[\em Exclusive shared control.] Different participants may control the same
INCO at the same moment in time and know that no other participant in control. In that case
control is shared (by a group)  but it is exclusive (for the group) as well. 
\item[\em Conjectural exclusive control.] A participant cannot be completely sure of its exclusive 
control, but it can conjecture that its control is exclusive.
\item[\em Conjectural exclusive shared control.] A participant cannot be completely sure of its exclusive 
shared control, but it can conjecture that its shared control is exclusive.\footnote{%
Conjectural exclusive shared control is a very plausible case if high amounts are at stake.}
\item[\em Legitimate acquisition of access versus illegitimate acquisition of access.] That an
INCO cannot be stolen or owned is irrelevant to the question as to whether an act of acquisition of control
of an INCO has been legal or illegal. Moreover, tracking transitions in an EXIM can be used to prove that
acquisitions were made in illegal ways, thus leading to prosecution and penalties, this of course dependent
of the legal system at hand.\footnote{%
A perhaps problematic comparison can be made with fatherhood. A man may illegally acquire the role of being
a father to a child. That does not change his status as a father to the child but it may bring with it, accusations, 
prosecution, conviction, and subsequent punishment, connected with the way in which his fatherhood 
came into existence in that particular case. The biological fact of being a father may be instrumental 
for this legal process to the extent that
without the biological information no accusation would be in place, 
and at the same time the legal process cannot undo the biological facts.}
\end{description} 

\subsection{P2P EXIMs and P2P EXINMs}
A special case arises if an EXI(N)M is based upon a peer-to-peer network. 
A peer-to-peer  EXI(N)M (P2P-EXI(N)M) is an  EXI(N)M with a particular structure and 
underlying mechanics. 
Consider a given EXI(N)M say EXI(N)M$_{h}$. 
It is a P2P-EXIM if the following conditions apply:
\begin{enumerate}
\item The participant base of EXI(N)M$_{h}$ contains a core of so-called peers of a P2P network. Peers operate
through interaction with their clients (that dedicated programs) on their computing devices.  These clients interact with the network, usually on the basis of a general purpose network such as the Internet.
\item Peers have equal voting right about the
future development of EXI(N)M$_{h}$. Such voting right may be weighted by the computing resources that
they have invested in maintaining a proper order in the EXI(N)M$_{h}$ P2P network.
\item Peers need to support the P2P network integrity by performing useful computational work. 
Peers need to participate in voting rounds needed for the disambiguation of the log files of EXIM$_{h}$.
\item Actions of peers may be honest or dishonest. Peers showing dishonest behavior are considered
dishonest themselves. 
\item EXI(N)M$_{h}$ has ways to ensure that honest peers are rewarded for their contribution to the 
integrity of the system, while dishonest peers are likely to be left without reward or even 
with some form  of penalty.
\item Honest peers are assumed to constitute a majority when voting about the evolution of EXIM$_{h}$ must be performed.
Therefore honest peers are supposed to determine the long term evolution of the relevant
P2P network and EXI(N)M (that is EXI(N)M$_{h}$) that it realizes.
\end{enumerate}

From now on we will simplify the notation and for the rest of the paper we speak of EXIMs only, even when dealing
with issues that work for EXINMs as well. We will speak of EXINMs in cases where the near-money is not a money.
 
\subsubsection{NA--compliant P2P EXI(N)Ms}
A subclass of the P2P EXI(N)Ms consists of those EXI(N)Ms that are based on an NA--compliant
P2P network. Roughly speaking that additional requirements adds the division of labour between ordinary
users and miners to the non-existence of a single point of failure which is a primary virtue of P2P networks, and
by inheritance also of P2P EXI(N)Ms.

\subsubsection{OSS NA--compliant P2P EXI(N)Ms}
OSS NA--compliant P2P EXI(N)Ms are those NA--compliant P2P EXI(N)Ms for which the evolution of software clients
is driven by an opens source community and for which state of the art clients (both for transaction  and mining) can
be obtained from an open source repository under an adequate open source license.

The current Bitcoin code base is reportedly OSS, thus Bitcoin can be classified as an OOS NA 
compliant P2P TIM if one regards the BTC as a unit of money and as an OSS P2P NA--compliant TINM if one prefers 
not to view the BTC as a unit of money. I may be the case that more than half of Bitcoin clients in use are not OSS software
products. For being an OSS system it is not essential that a majority of users employs an OSS client.

\subsection{Justifications for a positive valuation of INCOs of an EXIM}
A major worry that the perspective of the use of an EXIM creates for prospective users is that there is
no justification for assigning any value to an INCO. The value of an INCO expresses 
trust through information. The trust in a (hypothetical) EXIM, say EXIM$_{h}$, refers to several aspects and parties:
\begin{enumerate}
\item The trust that  producers of software clients are competent, honest and stable.\footnote{%
In case open standards have been fully developed and correctness proofs are doable trust in software
producers can be replaced by trust in proof checking agencies, assuming that software producers must
deliver their products inclusive a formal verification that can be checked by an independent party.}
For an OSS P2P EXIM trust is needed in the quality and stability of the open source 
software community which is leading the development of preferred clients.
\item  The trust that a participant may expect that exclusive control will not be compromised, that is
exclusive control will not degrade  into shared control with out the participant's consent.
\item The trust that exclusive shared control can
only be changed  with the consent of all members of the controlling group.
\item The trust that the participant base of EXIM$_{h}$ will be stable or growing.
\item The trust that the ICT (cryptography, protocols, etc.) employed will be of industrial 
strength many years to come.
\item The trust that it is unlikely that a new and competitive  or simply attractive
EXIM (say EXIM$_{h}^{\prime}$)  will force participants out of EXIM$_{h}$.
\item The trust that no participant can manipulate, either intentionally or accidentally, 
the overall quantity of INCOs in circulation in such a way that inflation undermines the value of INCOs.
\item In a period where classical non-informational monies are still dominant: the trust that EXIM$_{h}$
is a leading technology and that sooner or later non-informational monies will give way 
to informational monies, and that EXIMs will become dominant informational monies.
\end{enumerate}

None of these arguments prove that an INCO has some minimum value or that any object or service has
a counter value in INCOs. An EXIM merely provides an artificial bookkeeping of scarcity which 
can be used in combination with real bookkeepings of scarcity to maintain a perfect balance 
between holdings and obligations for a number of participants in parallel.

\subsection{Three dimensions of freedom for NA--compliant monies}
Summarizing the many options for variation that have been listed thus far the following picture emerges. Three
important dimensions of variation for NA--compliant monies can be distinguished.
\begin{description}
\item[\em Functionality.] Evolution of each realization will lead to more features being included. Below some
threshold the system incorporates a near-money rather than a money.
\item[\em Rules of engagement.] This is the dimension of autonomy and privacy. Many packages of rules of 
engagement can be tried out, with EXI(N)Ms as an extreme case of maximizing both user autonomy and user privacy.
\item[\em Mining process architecture.] Mining based on proof of work can be organized in many different ways and scaling
of an NA--compliant TIM, TINM, EXIM or EXINM to higher transaction volumes will require an evolution of
mining process architectures.
\end{description}

\section{Some remarks on Bitcoin}
\label{Getting-started}
The story of Bitcoin's coming into public existence is told in many papers.\footnote{%
For a history of Bitcoin including the development of mining technology  we mention~\cite{Taylor2013}. 
A thorough investigation to the social structure of the community responsible for Bitcoin can be found 
in~\cite{TeiglandYL2013}. Four important conclusions of that work are: (i) at least five key members of the 
Bitcoin Forum have been able to stay anonymous (at least until early 2013), (ii) Bitcoin is the 
outcome of a distributed group effort, and not merely the natural product/outcome of a single anonymous open software release,
 (iii) the Bitcoin community is self-organizing and to some extent self-healing,  and (iv) the Bitcoin Foundation
 operates according to its own principles.}
 We will be very brief about that and
limit ourselves to the reproduction of a single quote taken from~\cite{FaiolaF2013}:
\begin{quote} ``In January 2009, Satoshi Nakamoto unveiled bitcoin to a mailing list of computer super-geniuses. 
He, or she, or them--Nakamoto is a pseudonym for a programming whiz, or whizzes, whose 
identity remains one of the great mysteries of hackerdom Ñ effectively put a free 
software program on the Internet and invited the group to form a network of bitcoin Òminers.Ó 
They could excavate bitcoins by using computers to solve complex mathematical puzzles, 
with units of the cybercurrency seen as a reward for the electricity spent on the algorithms.''
\end{quote}

About Bitcoin we wish to state the some preliminary observations. 
We assume that readers have acquaintance with ``popular accounts'' of Bitcoin of which by now hundreds 
of versions can be found on websites and in blogs, in particular we will not repeat the common basics of 
transactions with inputs and outputs, the role of public key cryptography for transaction signing,
mining with diminishing returns from new BTC creation and increasing returns from transaction fees, 
hash rates, and blockchain growing.\footnote{%
An introduction to Bitcoin can be found in~\cite{Drainville2012}  and
information on Bitcoin clients is presented in~\cite{Skudnov2012}. An extensive survey of digital currencies including
Bitcoin is found in~\cite{Herpel2011}.}

\subsection{What Bitcoin is}
Bitcoin stands for a system,\footnote{%
We will identify Bitcoin as a system with the family of programs for clients as
located and maintained on \url{http://sourceforge.net/projects/Bitcoin/} initiated early 2009 
by the anonymous author the of the so-called Bitcoin white paper~\cite{Nakamoto2008} following
an early proposal for so-called b-money in~\cite{Dai1998}.
This definition of Bitcoin is
problematic in that it gives more prominence to this particular software than strictly needed.
Implicitly Bitcoin determines a protocol that can freely be used by anyone, and 
for that reason anyone can
participate in the Bitcoin P2P network with the support of a client of his or her own making
provided it complies with the protocol. As its stands now, the developers of Bitcoin (see
\url{https://Bitcoinfoundation.org/} for a survey of persons and their roles) 
 feel responsible for the sound working of the P2P network and act accordingly when problems
 arise (e.g. the treat of a hard fork at March 12, 2013 (see \url{http://en.wikipedia.org/wiki/Bitcoin} 
 for a brief explanation of that episode.)}
consisting of a special brand of P2P network, as well as for
the items circulating in that system, and more specifically for the unit of value that is 
used.\footnote{%
In~\cite{Nakamoto2008} no mention of named units is made and any positive rational that
occurs in a so-called verified transaction as an input or an output seems to qualify as a coin.
In subsequent documentation the unit is named BTC (unclear is to what extent BTC abbreviates Bitcoin), 
an estimate is made that no more that 2.1 10$^{6}$ will ever be (digitally) coined and the convention 
emerges that each BTC can be subdivided in to 10$^{8}$ so-called Satoshis.}
 BTC is the unit, and following~\cite{Nakamoto2008} positive rational quantities of BTC are coins,
and the system is explained in terms of a network and clients with various functionalities.
Bitcoin is a realization of the Nakamoto architecture that we have specified in Section~\ref{NakArch}
above.

\subsection{About the name}
The name Bitcoin suggests the presence of coins in the system. That is questionable, and in our  view
Bitcash\footnote{%
The name Bitcash has been claimed by several companies, research groups, and individual bloggers
 already, for instance see
\url{http://www.sutocorp.com/japanwebmoney/what-is-bit-cash} and \url{http://bitcash.cz/}.}
would have been technically a more appropriate name where cash is understood in a 
limited sense including only numbered banknotes with the numbers taken into 
account during each transaction. Stated differently, precisely coins and their capability of being handed over 
anonymously are missing from Bitcoin which turns it into a reduced product set finance (RPSF) in the
terminology of~\cite{BM2011}.

\subsection{Clients for Bitcoin}
Currently there are two types of software clients for Bitcoin users: 
transaction clients (also called users), 
which are the tools for Bitcoin users supporting 
the value exchange process,
and mining clients (also called miners) which participate in competitive block chain validation. 
Mining clients receive two kinds of rewards upon
winning a mining competition (at the moment such world-wide competitions take place between 5 and 10 times per hour).

The competition is about the construction of a block of successive transactions that show a perfect match with 
all blocks with earlier transactions dating back tot the first so-called genesis block that was mined in 2009. 
Rewards upon winning the proof of work competition with other miners consist of:
(i) proper mining rewards, a fixed but steadily decreasing number of BTC, a number that asymptotically approximates 0 and from 
about 2040 mining is specified to give no further returns, and (ii) toll rewards, collected by taking fees from
 each transaction included in the block being validated and promoted after proof of work as a candidate in the proof of work competition,
Bitcoin clients share a so-called Bitcoin protocol which concerns the following matters:
\begin{enumerate}
\item syntax of a transaction description (production by transaction client requires ECDSA key pair generation),
\item syntax of a block with proof of work, (checking by transaction client and mining client 
requires SHA-256 hashing and ECDSA digital signature checking),
\item syntax of a block chain, (checking by both types of clients requires the same cryptographic technique),
\item (specification of) an algorithm (to be performed by a miner) that determines a new blockchain from potential parts 
(this algorithm embodies diminishing mining returns), fee collection, 
\item (specification of) a blockchain reader (effectuated by transaction clients, determines priority between 
incoming candidates for the validation 
competition and participates in the voting in a predetermined way always giving priority to a blockchain of 
highest so-called difficulty).
\end{enumerate}
For transaction clients a further separation is useful: transaction creating clients, block chain reading clients, and 
transaction posting clients.

\subsection{Multiplayer game for cryptanalyzing SHA--256} 
If the hashing function SHA--256 is  broken, that is successfully cryptanalyzed, in some distant or not so distant future, 
that  may 	well happen as a consequence of its massive use in the proof of work component of Bitcoin mining. 
	
	One might even hold that Bitcoin merely constitutes a multiplayer 
	game (an applied game or alternatively a serious game in fact) that has been designed
	with the primary objective to successfully
	cryptanalyze SHA--256.\footnote{%
	As long as the identity of the Bitcoin designer(s) remains unknown
	it is hard to confirm that his/her/their objectives as stated in~\cite{Nakamoto2008} coincide with 
	``true" objectives.}
\subsection{Legal status, and where it stands}
The legal status of Bitcoin has become a complex and specialized topic already, which we will
cannot pay much attention to in this paper.

The question whether or not a quantity of $q$ BTC represents money
is far from obvious. One's philosophical judgement about the mater need not coincide with 
local jurisdiction about it. Such discrepancies may be a cause of problems.
In~\cite{SorgeK2012} it is clarified that within German Law Bitcoin does not qualify as ``Digitale W\"{a}hrung''
because that qualification requires legal tender status, and Bitcoin also fails to qualify as ``E-Geld'' because it carries
no claims against an emitting agent. In Germany BTCs are an official money of account, however, 
since September 2011. That status poses restrictions on those who offer goods and services in return for Bitcoin.

Bitcoin might constitute a new phase in the history of identity management 
(see e.g.~\cite{Leeuw2007,Higgs2007,LeeuwB2007})
where abstract numbers (public keys) take priority over personal identities and voting rights are given 
on the basis of proven processing power rather than on the basis of a human headcount, or an a
approximation thereof via IP numbers.

Since 2009 Bitcoin has become an international movement attracting wide attention.\footnote{%
Because Bitcoin is only one of many novel informational currencies it requires an explanation why 
Bitcoin is so attractive. Before, in the 1990-ies, David Chaum's informational money 
design named DigiCash has attracted quite some attention, and
therefore the media-hype surrounding Bitcoin is not really new. DigiCash, however, never flourished, however, and
as a company it failed and terminated in 1998. Bitcoin outperforms expectations systematically, and that
is helpful for building a hype, which by itself justifies higher expectations and so on.}
For instance in an anthropological assessment (\cite{MaurerNS2013}) the case is made that 
Bitcoin embodies so-called digital metallism. The Bitcoin terminology reminds 
of metallic monies through the terminology of mining. At the same time Bitcoin miners must 
be very materialistic because without significant investments in computer 
hardware and software their attempts to dig up new Bitcoins will be fruitless.\footnote{%
In~\cite{MaurerNS2013} the software (code) and cryptography are identified, both in 
combination referred to as code. This approach seems false to us, a Bitcoin 
user must have confidence in the cryptography and may doubt the validity of the used 
client software at the same time.}

\subsection{Assessment as a money: functionality and ethcis}
\label{AssessmentofB} We will make an attempt to provide a very preliminary 
assessment of Bitcoin in terms of the characteristics and 
other aspects of monies that have been listed in Section~\ref{Wim} above. We will provide two ratings in the 
range (very good, good, adequate, weak, inadequate/absent, unclear), the first 
rating assessing Bitcoin's current performance (CP), 
the second one assessing its ultimate potential (UP) under the current rules of the game).
\begin{enumerate}
\item Barter catalyst: CP = adequate (technology for application specific clients and wallets still less developed, transaction
validation is still too slow), UP = good 
(very low transaction price, and very simple actions required because no IBAN, names, goods/services are to be provided).
\item Unit of account: CP = inadequate (very unstable), UP = poor (no protection against deflation),
\item Store of value: CP = good (the past performance was very good at the time of this writing), UP = very good, (the best scenario
for Bitcoin is amazingly good),
\item Expression of freedom: CP = very good, UP = very good,
\item Functional communication protocol: CP = weak, UP = unclear (in item~\ref{ethics} 
below we will return to this issue),
\item Enabler of human rights provision: CP = weak (but some forms of participant autonomy are 
taken care of much better than in conventional monies including their informational incarnations), UP = unclear,
\item Instrument of internal governance: CP = weak, UP = inadequate (our guess),
\item Instrument of external governance: CP = absent (inapplicable as Bitcoin denies the 
significance of geographic qualification of
boundaries between internal and external), UP = absent.
\end{enumerate}

And here is a preliminary and very incomplete assessment of Bitcoin in ethical terms:
\begin{enumerate}
\item The computational cost of mining/transaction verification measured in terms of 
energy may be hard to defend: with more mutual trust one
may be able to design an effective transaction verification (and double-spending prevention) system against
 lower energy costs. (See~\cite{Herrmann2012} for a case study on performing double-spending attacks as 
 well as prevention of such attacks.)
\item The Bitcoin design seems not have taken into account the possibility that
Bitcoin succeeds and becomes so important to many people that the need to manage drastic fluctuations 
emerges on ethical grounds. Bitcoin ethics is relatively easy as long as Bitcoin is marginal and its proponents 
must fight for its survival. Nevertheless, we can imagine a situation where the OSS community develops the vision that
it needs to take its own responsibility in bringing the Bitcoin system down because some of 
the consequences of its victory cannot be morally defended. 
	\begin{itemize}
	\item In particular we can imagine that the lack of anonymity provided
	by Bitcoin eventually poses a privacy problem against which the open source community will need to take a stand.
	\item If a sub community must survive temporary isolation from the network at large it is unclear to which 
	extent they can use Bitcoin
	as a local money and have their part of the blockchain integrated in the main part afterwards. 
	Depending on Bitcoin in adverse circumstances may be imposible.
	\item Even more problematic is the situation during a sustained breakdown of electric power supply. 
	\end{itemize}
Give that each of these cases constitute real risks it is essential that a success of Bitcoin doe not come at the expense of the availability of a robust money that can function in a range of adverse circumstances.
\item Early adopters of Bitcoin have gained a very significant advantage over later newcomers to Bitcoin. 
Given the deflationary bias
of Bitcoin that seems to be rather unfair. Technically Bitcoin may not be a Ponzi scheme because it may properly 
function with a stable 
user base and Bitcoin is not necessarily dependent on the influx of new external money from new users,\footnote{%
The Bitcoin store of value mechanism includes a Ponzi scheme, but openly so to speak, while the efficiency and productivity
of Bitcoin as an exchange mechanism is real and has an economic value that can be expressed in the 
size and activity of its participant base just as that can be done with other social media.}
but the early adopter's advantage is so pronounced that in later years one may be inclined to speak of a Bitcoin 
scheme (or a Nakamoto scheme) instead.\footnote{%
These issues have been highlighted in convincing detail in a remarkable 
survey: \url{http://www.gizmag.com/Bitcoin-creation-value-overview/26325/}.}

This aspect of Bitcoin we consider rather worrying and morally problematic.\footnote{%
Having this worry may signal a generational problem in that we are perhaps unaware of the real
drivers of success in modern IT. 
It may have become a social norm that founders of a 
successful new social medium, or of an iconic part of the  IT industry,  must become very rich for the medium,
or the industrial designs, to attract massive public attention. If that
is true then the Bitcoin episode indicates that even anonymous and hypothesized wealth can play that role.
Moreover, if founders becoming rich is a necessary condition for success, rather than a mere consequence of it, the
earning model of Bitcoin constitutes a respectable innovation. And many will agree that if Bitcoin rises to prominence
those who got it going deserve huge profits.}

\item This can be held against a worry about excessive early adopter advantage: 
those who eventually end up with many BTCs that were obtained cheaply 
(at the time of writing this Paragraph the increase in value was with a factor 10,000 when compared with the value
around July 14, 2010, just before Mt.Gox was established as the first Bitcoin exchange service) have shown faith in
the system in spite of spectacular fluctuations. Not selling during early peaks in Bitcoin rate, and thereby repeatedly 
taking the risk of not gaining significant profits constitutes an important contribution to its rise to prominence just as well.

\item The idea that most BTCs may still be in the hands of a few individuals
is progressively hard to swallow if Bitcoin becomes more mainstream. The start of the system gives so much 
advantage to early adopters that
those who buy BTCs for big money (in terms of Euros) must feel let down. Having made the point that ``this can be done'',
a fair restart seems to be necessary at some stage.

\item There is a risk at any time that a coalition of miners discriminates agains the transfers of 
specific classes of agents. 
\item There is always a risk that mining becomes 
monopolized in the same way as all approaches to the design of social media have become thus far. 
Miners can work at zero cost for participants willing to
cooperate with their media ambitions and so on. The system is 
so open and public that mining agents can interact with those
whose transactions they have verified. A strong miner may 
point out that mining may be performed without fee only initially. 
\item In circumstances that Bitcoin would become a cause of trouble: who is responsible for it? All programmers plus all
miners plus all users plus all hoarders (inactive users)? 
\end{enumerate}

\subsection{Documentation}
\label{NoOfItems} 
Currently there is a lack of published specifications and descriptions 
of the intended working as well as the actual of up-to-date Bitcoin clients. 
Lacking a convincing formal specification, 
which is binding for all participants of the ``Bitcoin game'', it is unclear to what extent the 
developer team can (or should, or intends to) control its future evolution.\footnote{%
The comparison with other open source software engineering
projects is only of limited value once the existing Bitcoin volume is becoming understood as a 
true distributed store of value upon which participants intend to rely in an economic sense.} 
The open source approach implies that technically stronger alternatives to Bitcoin may be 
developed while making full use of existing Bitcoin software, and by members of the original 
development team, and without showing any commitment to whatever form of maintenance 
of the market capitalization that has been achieved within the current Bitcoin user community. 
Viewing the Bitcoin P2P network as a software product, or even as a social medium, the 
open source ideology must always be open to and supportive of 
the development of competing products by 
incremental improvements of the existing ones.\footnote{%
The open source software and systems (OSS) movement is committed to obtaining
best possible software products, rather than to see its own offspring to victory in spite
technical difficulties that could have been solved in a really free market.}
On the long run there seems to be an unavoidable internal 
contradiction between the Bitcoin project as an open source software effort and 
the ``Bitcoin project'' as dedicated approach towards opening up and restructuring existing
financial and monetary technologies.

\section{Information security aspects of EXIMs}
\label{Security}
In a community where gold represents money agents can sometimes 
display their wealth and store it in a way accessible
for all other agents without security measures, while relying for the protection of their property 
on the social contstruct of ownership only. With an
EXIM there is no reliance on an enforceable social contract while making one's informational wealth (in 
terms of control rather than ownership) publicly visible  is certainly  possible.

For an EXIM security is not merely a matter of implementation but it is the heart of the matter. Security determines the
only mechanism by which value can be created, transferred, and stored. It follows that maintaining some abstract 
view of security is inescapable for every user of an EXIM.\footnote{%
This conclusion applies in practice to Bitcoin users as well, but assuming Bitcoin is a TIM rather than an EXIM,
the need to understand its security model (and its implementation) plays  a pragmatic role 
w.r.t. Bitcoin instead of the principled role that it plays w.r.t. an EXIM.}

Providing a comprehensive introduction to computer security is not the purpose of this Section, rather 
highlighting a number of aspects that might be characteristic of the security issues surrounding an EXIM. We will
first list some assumptions that we will make without further substantiation.

\subsection{Classifying viewpoints about information security and computer security}
An important objective of this paper is to formulate some conceptual premises on which security
considerations can be based. Unfortunately the various explanations and listings below are less structured
then one might hope. We hold that in order to formulate and understand a particular viewpoint
on security it is needed that one maintains a number of technical assertions embedded in a framework
of informal assumptions. We are interested in the informal assumptions that come into play. To begin with here is
a very general classification of viewpoints (referred to as assumptions) concerning security.
\begin{description}
\item[\em Constructive assumptions about the use of tools and methods.]
Many assumptions explain the virtue of using particular tools and methods. For instance:
	\begin{itemize}
	\item It is useful to install all security updates of an OS as soon as possible.
	\item Physical security prescribes  care with the use of radio signal emitting equipment.
	\item Passwords must have at least 10 characters of different kind and must look rather random.
	\item Passwords must be changed each year.
	\end{itemize}

\item[\em Assumptions that something won't go wrong.] Often one will argue that whenever a certain
kind of security problem is present the precautions that have been taken were so effective that the presence of the
problem won't go unnoticed.
	\begin{itemize}
	\item If you don't speak loudly and see nobody around, then what you say won't be overheard by any third party.
	\item If you are in a random location and you have a text on your screen and nobody who might be looking
	at you can be seen, you have exclusive access to the information on the screen.
	\end{itemize}
	
\item[\em Assumptions that some form of security cannot be guaranteed in a certain way.]  For instance:
	\begin{itemize}
	\item Without assuming the presence of a valid security model, as well as its correct implementation,
	 the security of a session with a  general purpose computer cannot be claimed.
	 \item Without knowing a reduction to a problem of known (or widely assumed and agreed) 
	 computational complexity a cryptographic method
	 is to be considered less reliable.
	\end{itemize}
\end{description}

\subsection{Assumptions about computer security}
\label{assumptions}
We believe that no account of information and computer security can do without general assumptions. 
But people differ
on which assumptions they will subscribe to. Each explanation of an EXIM must unavoidably 
deal with three forms of security:  information security, physical security, and with computer security.
We subscribe to the following assumptions concerning security, as well as to the outline of a 
secure EXIM handling process model ({\em security by forward physical separation}) 
that emerges from these assumptions.
\begin{enumerate}
\item Agents are supposed to have an intuitive understanding of physical security, 
in spite of the fact
that physical security is very hard to achieve. An agent $P$ should not be supposed to have any reliable
intuition concerning the notion of information security, and in particular not about computer security. No
tradition of installing firewalls, performing security updates of one's OS, changing passwords, and making
password ``difficult'' generates any understanding of what a reliable and potentially provably correct security
model might amount to.

\item Suppose an agent $P$ acting as a normal but competent and experienced user, 
makes use of a computer $C$ with internet connectivity, then $P$ cannot be sure that nobody
is looking over his/her shoulder in a digital way, and for that reason $P$ cannot be sure that what 
is displayed  on a screen visualizing memory content of $C$ is actually private.

\item In particular knowledge of so-called security models meant to explain how an operating system 
handles security and access to resources allows no more than an initial 
understanding of how a system might work,
similar to introductory explanations of the chemical structure DNA which don't explain the mechanics of genes, but
merely uncover a natural biotechnology on the basis of which genes may exist.

\item Similarly, the hypothetical computer user $P$ cannot know for sure that data stored in his machine $C$
have been completely removed, and connecting $C$ to the internet introduces a risk that all data that
$P$ has previously  handled on $C$ become accessible to an unknown intruder.

\item In order to overcome the doubts concerning security as stated in the previous items $P$ needs to 
have a reliable security model at hand together with a verification
of its crucial properties and a verification of the entire implementation down to the operating system and the hardware. 
In practical terms this burden cannot be placed on the shoulders of a solitary competent computer user, or of any team
operating in a conventional manner, however professional the team may be.

\item {\bf Security by forward physical separation.} In an NA--compliant EXIM, 
all private keys giving access to large amounts of information money
must always reside on devices that will not, 
in any later phase of their operation be connected to the internet. 

Here is an account of a secure EXIM usage model or formulated differently 
a suggested practice for performing EXIM transactions. This 
account should not be taken literally, it is meant to express by way of example which steps may be needed in order to
arrive at a security policy which may be reliably grounded in $P$'s potentially valid security intuitions. The assumption 
that this kind of practice is secure or can be made secure, is itself one of the assumptions that we are listing.
	\begin{enumerate}
	\item After installing $C$'s OS and utilities it may be connected to the internet in order to download data 
	relevant for the P2P network status related to the EXIM at hand. After these preparations which must
	precede each (uninterrupted sequence of) transactions $C$ must be disconnected from the internet.
	\item After any production of a key pair (on a free standing computer $C$), 
	EXIM participant $P$ may proceed as follows:
		\begin{enumerate}
		\item compute with the help of $C$ the signed (with the secret key just computed) data needed for 
		a transaction thus producing an INIT $\alpha$, 
		\item and export that INIT $\alpha$ in combination with the public
		key (but without the secret key), together (as a cluster INIT) in a physically secure way (a notion intuitively 
		grasped and effectively guaranteed by $P$) from $C$ to an internet connected machine 
		$C^{\prime}$, also operated by $P$ (but not necessarily exclusively operated by $P$), 
		\item when operating $C^{\prime}$ security hardly matters (as $C^{\prime}$ contains no confidential information)
		as long as $P$ gets assurance that transactions are posted (no DDoS attack blocks
		$C^{\prime}$ from posting), and as long as $P$ can be sure that received feedback is completely shown.
		\end{enumerate}
		\item Now $P$ must choose between the following two options:
		\begin{enumerate}
		\item$C$ is physically destroyed. (It may suffice to clean $C$, that is to remove all 
		software on $C$ and reinstall everything
		from scratch, if it is sure that the hardware is memoryless after the removal, a feature which can be guaranteed
		by $C$'s construction. It cannot be much cheaper than that.)
		\item $P$ takes the decision that $C$ will be used for some time to come 
		for storing private key information. In that case $C$ can be used
		for that purpose only and should not be reconnected to the internet, or be operated by any other 
		untrusted user $U$  		before it has been cleaned, and moreover $P$
		is in charge of the physical security of $C$ until it has been cleaned or destroyed.
		\end{enumerate}
	\item In both cases just listed, $P$ may or may not preserve the physical carrier that has 
		been used to carry the signed cluster INIT mentioned before and
		stores that physical item securely (which may in turn depend on computer use and so on).
		
	This protocol is not entirely obvious and if many transactions are performed the policy involved needs to be 
	designed in advance, written down on paper, and preferably validated by an independent security consultant.
	\item Subsequently $C^{\prime}$ is used by $P$ to take care of posting the transaction
	on the P2P network and finding out whether that transaction is accepted and so on.
	\end{enumerate}
	
\item From the above it follows that $P$ can in principle operate an NA--compliant EXIM without the use of firewalls,
passwords, and similar classics in computer security, provided that $P$ has an unshaken faith in public key 
cryptography (especially in digital signatures) and in $P$'s own competence of maintaining physical security. 
Conceivably a decision
not to make use of any other password than one's own publicly known name increases the 
security of EXIM handling, provided the consequences of that decision are taken seriously, which implies that physical security 
must be brought to the forefront.

What is secure for an EXIM or EXINM  is secure for a TIM or TINM as well. As we have classified Bitcoin as an NA--compliant
P2P TIM the protocol just mentioned is also effective for a Bitcoin user. We will return to that matter below.
\item In order to trust the use of cryptography $P$ must be sure that $C$ performs the (known and public)
cryptographic algorithms correctly. This is not an easy task but is does not require the understanding of anything 
beyond sequential computing. Multi-threading enters the picture once cryptographic algorithms require 
advanced concurrency that must be mediated via a operating system. 

The use of computers, in our case  $C$ is only necessary
because agent $P$ can not perform all required computations by hand in time. The need to deal with computational complexity
may force $P$ to steadily improve the computer technology used for the role of $C$.)

\item Faith in public key cryptography is a matter of mathematics and probability, not a matter of computer security.
Confusion of these issues is problematic. In order to understand the connection between an EXIM and public 
key cryptography some meta-theory is needed. Below we develop the notion of ``Conjectural almost pseudomonopresence''
as an example of how such a meta-theory might look like.
\end{enumerate}

\subsection{Informational item types: monopresence, pseudomomopresence, and conjectural pseudomonopresence}
\label{INITtypes}
The concept of information has been analyzed comprehensively and in painstaking detail in 
\cite{Floridi2011} but it is difficult to obtain from that account a notion of information that supports an EXIM. 
A notion of information is needed that  conforms with the use to be made of it in an EXIM. 

Below we will provide a slow built-up of notions of informational item and informational item type
that seem to underly the intuition of an agent being in control of some amount in an EXIM.
\begin{description}
\item{\em Abstract informational item (ABINIT).} An ABINIT is a finite
mathematical quantity $q$, for instance a rational number or a tuple of natural 
numbers. 
\item{\em Abstract informational item type.} An ABINIT $q$ is usually an element of some type $T$. $T$ is an ABINIT type.
In the setting of an EXIM ABINIT types are finite sets, for instance bit sequences of a 
fixed length. 
\item{\em Identification of ABINITs with finite bit sequences.} In a first version of a theory of information items it is 
reasonable to identify the union of all ABINIT types with the infinite set NAT of natural numbers. When needed we will assume 
that some natural encoding of an ABINIT type into NAT is available.
\item{\em Informational item.} An INIT (a ``real" piece of information) is a physical representation of an
ABINIT. An INIT may exist in different forms for which further classification is useful:
	\begin{itemize} 
	\item static and immobile, e.g. an inscription on a building, or a data item stored in a mainframe data 
	base on a fixed location.
	\item instrumented static and mobile, e.g. part of a printed text (say a book), or data stored in a mobile phone, or in a laptop.
	\item human agent based static and mobile: located in and retrievable only from the memory of a living human agent. 
	\item dynamic, e.g. in the form of a traveling data packet, or an item briefly visualized and broadcasted on a TV screen,
	or a text being rendered on a single computer screen.
	\end{itemize}
This classification is rather imprecise. Static INITs may have different degrees of stability, robustness against
external impact, and expected life-time. Mobility can be distinguished as physical/mechanical mobility, information 
technology mediated mobility, and human interaction based mobility.

\item{\em Representational chain.}
A physical representation, mediated by a representation scheme $r$, of an ABINIT $q$ is consists of these components:
\begin{enumerate}
\item A syntactic representation of $q$ as a term $t = t_q$ in some appropriate syntax $S_r$.
\item An encoding $E_r$ of $t$ in a bit sequence $s= E_r(s_t)$. 
\item A physical form of $s$, say $p=p^s$, (said to represent $s$).
\item A measurement protocol $P_r$ which allows 
reading off $s$ (that is deriving or inferring  $t$ from $s$) by known methods of observation. 

\end{enumerate}
The list $(p, P_r, s,E_r, t,S_r, q)$ is
a representational chain for $q$. Given the concept of an informational chain, we have the following conventions:
\begin{itemize}
\item We will write $q= \alpha_r(p^s)$. Here $\alpha_r$ extracts an ABINIT from an INIT
under the assumption that representation scheme $r$ is used.
\item For an INIT $p =p^s$ to exist a representational chain of which it constitutes the head 
must exist as well.\footnote{%
A significant, but seemingly unavoidable, complication for this definition is that parts of the representational 
chain, in particular, $P_r, E_r$, and $S_r$ may exist in the form of evolving conventions in the minds 
collaborating groups of of human agents only. This leaves the concept of an INIT vulnerable to questions about
independence of human agency, and about persistence when parts of a representational chain are unavailable or are temporarily unavailable. It also creates INITs that are INITs for some but not for all. This complication leaves one also
without an answer on very plausible questions of the form: ``here are 10.000 INITs with respect to 
representational method $r$, so what is $r$?"}
\item $p=p^u$ and $q=q^v$ are representationally equivalent with respect to $r$ id $\alpha_r(p)=\alpha_(q)$.
\end{itemize}

\item{\em Informational item types.} INITs are a part of physical reality, and as such an INIT $p$ can be described. A
description in full detail will disclose what is known about the representational chain of which $q$ is the head. In practice it is important to be able to speak of classes of potential INITs. That leads to the notion of an INIT type which is best viewed as a
specification of an initial part of a representational chain. Given a representation scheme $r$ an INIT type $T^r_i$ 
specifies a collection of possible INITs for $r$ so that many different INITs $q_i$ may comply
with $T_i$ and the corresponding values of $\alpha_r(q_i)$ may also differ. 

Many different INIT types can be imagined, in some cases a type may specify a range of representation schemes rather
than a single one, or it may fix $P_r$ while leaving room for a range of $E_r$'s etc. Viewed from the perspective of 
``the INIT type of $q$", including $\alpha_r(q)$ as a component of an INIT type description is less plausible, while the more
plausible way to deal with an INIT type lies in answering questions like: 
\begin{itemize}
\item here is INIT $q$ of INIT type $T_i$, what ABINIT does it represent i.e. is what is $\alpha_r(q)$, 
\item given ABINIT type $T_a$, design (or select) an INIT type $T_i$ suitable for representing all elements of $T_a$, and
\item given ABINIT type $T_a$,  and representation schemes $r$ and $r^{\prime}$ with corresponding 
INIT types $T^r_i$,  $T^{r^{\prime}}_i$suitable for representing all elements of $T_a$, which of the two representation types
is most suitable for a certain application area.
\end{itemize}

\item{\em (Representational) presence of an abstract informational item.} 
An ABINIT $q$ has presence (that is ``is represented"), 
in the form of an INIT $p$ 
 if $q$ is the 
endpoint of some representational chain beginning with $q$. 
Representational presence is linked to place and time. Existence of $q$  is another way to speak about its representational 
presence (somewhere) at a given moment of time.
\begin{itemize}
\item We will assume  that the number of existing INITs is finite at each moment in time.
We propose to consider the Milky Way as that part of the universe to which such statements of existence
or of unique existence make reference.\footnote{%
A non-existing INIT difference from an ABINIT in that its existence (given a description of the corresponding representation) 
at a certain time and place may still be hypothetical, 
which is an uncertainty that cannot pertain to an ABINIT.}
\item  It follows that at each moment in time only a finite number of elements of NAT is present. Estimating the
size of the set of ``existing" natural numbers is a difficult matter, it clearly depends on conventions about 
representational chains.

\item Not only is the world finite, some upper bounds of its size are considered plausible.
Common ABINIT types such as bit strings of length 128 are so large that only a small fraction 
of their elements can be present simultaneously in our world (or perhaps even in the part of the universe
that may be accessible to mankind.
\item Natural numbers which are implicit, in the world, (e.g. the number of stars in the Milky Way above a certain mass),
are not considered present as INITs, through their mere implicit existence. Estimates of such numbers written in research
papers do qualify as INITs of course, quite independently of their scientific accuracy.
\end{itemize}
\item{\em (Pseudo)random generation.}  In the context of an EXIM an INIT $s$ is plausibly a relatively small, 
(say with 10,000 elements or less) bit 
sequence physically represented in time and space on some medium (which may include a human brain).
$s$ combines several parts which have been obtained via one or more of the following steps: 
(pseudo)random generation, input of transmitted information, application
of cryptographic operations  (hashing, signing, encryption) to bit sequences.
\item{\em Monopresence.} If a mathematical quantity $q$, that is an INIT, has presence via a single
representational chain only at some moment of time it is said to be monopresent at that moment. 
\item{\em Multipresence.} If a mathematical quantity $q$ is represented by different INITs 
$p^s_1,..., p^s_n$ simultaneously  then $q$ is multipresent. $n$ ford multipresence, for some natural number $n>1$ 
is common for public keys, names (of agents, objects, persons, and organizations), user names for computer systems, 
pricing data. Symmetric keys are likely to have no presence or double presence.
\item{\em Pseudomonopresence.} In case presence of an INIT is mediated by technology which necessarily
provides causally connected occurrences of that INIT, multipresence occurs, though in a limited fashion.
In this case we will speak of a variation of monopresence:
pseudomonopresence. 

An example of pseudomonopresence is found if a computer has an INIT in 
memory and on screen, or in memory and after it has been printed. We also speak of pseudonomopresence
if a backup has been made by an agent in control of an informational item. 

There are several 
complications with the notion of pseudomonopresence, asking for further refinements:
	\begin{description}
	\item{\em Conjectural monopresence.} It seems to be an assumption (conjecture) that 
	whoever randomly generates a sequence $s$ of 128 bits, resulting in an INIT $p^s$ can 
	be practically sure of the monopresence of the string $s$. Monopresense of $s$ is 
	a conjectural matter because collisions cannot be excluded for independent random generating mechanisms.
	\item{\em Conjectural pseudomonopresence.} It seems to be an assumption (conjecture) that 
	whoever randomly generates a sequence $s$ of 128 bits, resulting in an INIT $p^s$, and after having made
	one or more backups, can 
	be practically sure of the pseudomonopresence of the string $s$. Conjectural pseudomonopresence rather than
	``true" pseudomonopresence is the epistemic state of affairs that an agent will reach.
	\item{\em Trusted pseudomonopresence.} An agent who believes in the 
	pseudomonopresence of an INIT $i$ still
	needs to have trust in the fact that the different representations constituting $i$'s 
	multipresence are all
	causally connected to the same origin in a way that complies with the design of the technology
	that is being used.\footnote{%
	An agent copying a character string with pen and paper under the assumption of preserving trusted
	pseudomonopresence assumes not being monitored by a hidden camera and so on. 
	Not being compromised is a 
	matter of maintaining high standards of physical security, psychological security,  
	and computer security.}
	\item{\em Compromised pseudomonopresence.} 
	If another agent has obtained access to 
	at least one representation (or a significant part of it) of a multipresent and
	pseudomonopresent INIT the agent's conjectural pseudomonopresence of the INIT has been 
	compromised.
	\item{\em Accidentally violated pseudomonopresence.} If one or more 
	representations of the INIT after its supposedly original 
	generation are not causally connected to that process
	of generation (that is: against all statistical odds the allegedly newly created INIT
	existed somewhere else already, that is an equivalent INIT was in existence) then the
	pseudomonopresence of that INIT is said to have been accidentally violated.\footnote{%
	An agent $P$ creating a four digit numerical PIN code for a credit card knows in advance
	 that monopresence 	and pseudomonopresence of the 
	resulting INIT  $s$ are both extremely unlikely and probably simply impossible to achieve. 
	It is not even problematic
	(though useless) for another agent $Q$ to have a listing of all four digit codes in a private 
	database so that $Q$ has access to $s$ even before $P$ created its new representation(s). 
	Clearly the security 
	protection that $P$ introduces by its creation of new representations of $s$ cannot
	be explained in terms of presence, monopresence, pseudomonopresence, 
	or any variation of these notions.}	
	It is an implicit assumption that no agent will be able to misuse the accidental
	existence of violating
	representations of an INIT that was created originally under the assumption of 
	achieving pseudomonopresence.\footnote{%
	This assumption requires competence of the agent (say $P$) making it: competent use of 
	(pseudo)random generators, avoiding any connection between INIT content and other existing
	INITs that might have been either communicated or leaked to other agents in 
	such a way that they may link that INIT to $P$.}
	\item{\em Almost pseudomonopresence.} Each agent generating an INIT under the hypothesis of
	conjectural pseudomonopresence is aware that certainty cannot be achieved. In fact a slightly
	weaker assumption may be used:  almost pseudomonopresence: (i)
	pseudomonopresence has not been compromised, while (ii)  it may have been accidentally violated.
	\item{\em Conjectural almost pseudomonopresence.} Unfortunately it is impossible for an agent
	to be completely sure of
	almost pseudomonopresence in any actual context. As a consequence that fact must be 
	conjectured, leading to conjectural almost pseudomonopresence as the default that an
	agent maintains about an INIT that it has just created (or instructed a device to create, or for which the
	agent has created a backup) with the 	firm objective to obtain pseudomonopresence.
	\end{description}
COMMENT: Pseudonomopresence rather than monopresence is the normal intended state of affairs for secure keys
in computing, and conjectural pseudomonpresence is the normal intended epistemic state (state of knowledge) 
that goes along with it.

\item{\em Logical and physical access to information items.} Accessibility of INITs is relative to agents 
supposedly having that access. Taking an agent into account a distinction between logic access and physical access
must be made. Representational schemes specify forms of logical access and forms of unproblematic physical access
at the same time.
\begin{description}
\item{\em Immediate logical access to an information item.} We will speak of immediate logical 
access to an INIT $q$ in cases 
where a representation scheme $r$ can be assumed without further doubt or specification. Such assumptions 
may be non-trivial, and ad hoc. For instance seeing a written representation on a blackboard may not be helpful
in a room without light.
\item{\em Logical access to an information item.} An agent $P$ has access to an INIT  $q$ under representation scheme $r$
 if $P$ has a method, taken from a ``small'' set of  methods, which allows it to read off the INIT.  
Reading refers to any form of successive awareness, either through vision, 
through hearing, or through tactile or other senses. Logical access implies the availability to an agent 
of a transformation of $q$ to an INIT 
$q^{\prime}$ to which the agent has immediate access w.r.t. a representation scheme $r^{\prime}$ and such that
$\alpha_r(p)= \alpha_{r^{\prime}}(p^{\prime})$.

\item{\em Immediate physical access to an information item.} We will speak of immediate physical 
access to an INIT $q$ for an agent $A$ if (i) $A$  has immediate logical access to $q$ w.r.t. $r$, (ii) this particular form
of logical access as such, and without any additional complications or difficulties specifies a mechanical way for $A$
to inspect $q$, and (iii) upon generalizing immediate access of $A$ to a suitable INIT type $T_i$ for $q$, $A$ is able to
assess representational equivalence of two INITs (w.r.t. to $r$) by means of this method of inspection.

\item{\em Physical access to an information item.} Physical access to an INIT requires a distinction of two different cases.
In the simplest case we speak of direct physical access. The second case, however, applies to all forms of secure storage
if an INIT. 
\begin{description}
\item{\em Direct physical access.} We will speak of physical 
access to an INIT $q$ for an agent $A$ if the description of $A$'s logical access to $q$ also indicates an unproblematic 
method of physical access.
\item{\em Indirect physical access.} Indirect access of $A$ to INIT $q$ applies if $A$ can follow a representational chain with
head $q$ only after engaging in and additional access protocol. Carrying out that access protocol may require $A$ to
either have a physical key at hand in order to get at $q$'s physical location, or it may require $A$ to have direct physical 
access to some INIT $k$ that serves as an informational key for an informational storage of $q$.
\end{description}
COMMENT: Informational key (i.e. password protected) enabled indirect (non-immediate) access to an INIT $q$, possibly complemented with physical key enabled immediate access to a (safely stored and) representationally equivalent INIT, 
describes the way in which 
a secret key $k = \alpha_r(q)$ for a Bitcoin account $a$ is stored in a wallet, by an agent $A$ who considers itself in 
control of account $a$. $A$ will assume conjectural pseudomonopresence of $Q$ as well as of the informational key $u$
enables $A$'s access to $q$.

\end{description}
\item{\em Meaning of INITs.} The meaning of an INIT is a topic quite different from its representation. We will make two remarks only about INIT meaning.
\begin{description}
\item{\em Structured ABINITs} The meaning of an INIT is found as the meaning of the ABINIT it is supposed to represent.
Meaning is necessarily relative to a representational scheme $r$. An ABINIT requires subsequent parsing 
and decomposition into its constitutes parts.
\item{\em Roles instead of meaning.} The meaning of an information item first of all
lies in the roles played by its components. These roles often have technical names such as: quantity
of transaction, time, public key, private key (or secret key), nonce, hash, transaction data. 
Roles may be given implicitly
by way of aspects of a method for retrieving the information, or as parts of an information item.
\end{description}
\end{description}

Having developed this somehow unwieldy terminology it can now be stated that the 
security of INCO handling in an EXIM should primarily be based on participants ability to 
create INITs under the assumption of conjectural almost pseudomonopresence. That justifies the
assumption of being unique in having access to self created INITs at will.

\subsection{Achieving conjectural almost pseudomonopresence}
If one can safely assume that one's computer is secure under all circumstances
then the conjectural almost pseudomonopresence of a self generated key pair of
sufficient length follows from the default assumption that no other data transport or interaction 
takes place than what a user explicitly intends to achieve. 

This degree of security can be achieved along the lines spelled out above, with a strong reliance on
physical separation of devices. Classical information security constitutes the basis of this approach,
at the exclusion of traditional computer security.

The intuition of physical security, that nobody is watching so closely what someone is doing that
information is unintentionally leaking, seems to be reasonably clear. It may be hard to achieve, 
but at least it seems obvious what one intends to achieve. 

\subsection{Bitcoin can be surprisingly secure}
\label{Bcbss}
As we have mentioned already Bitcoin is classified as an OSS NA--compliant P2P TIM. That implies that we
may apply the preceding description of EXIM security to the special case of Bitcoin.
Suppose that Bitcoin is used by $P$ in the following manner:
	\begin{enumerate}
	\item $P$ does not care about mining and is happy to pay some fees when performing transactions. If $P$
	participates in mining as well that is done on completely independent and permanently separated equipment.
	\item The transaction client is split in three independent components, thus obtaining three clients: 
		\begin{itemize}
		\item User input client (incoming transactions, double-spending notification (optional), 
		validated block receiver (from miner), multiple block chain preference determination, 
		and blockchain checker),
		\item Transaction constructor client (requires handling of secret keys, block chain received from user input client before
		secret key is either generated or used),
		\item Transaction placement client (including transaction input from transaction constructor).
		\end{itemize}
	\item The transaction constructor client is generated each session from scratch and is executed on a machine
	playing the role of $C$ in the description of security by forward physical separation.
	\item The user input client and the transaction client are both running on a device (may be the same) 
	with permanent internet connection. 
	\item Secret keys generated by $P$ are always kept safely in store an only handled by clean instances of the
	transaction constructor client.
	\item A high level of physical security is observed.
	\end{enumerate}
In this setting $P$ can be assured that:
\begin{enumerate}
\item $P$ will not lose exclusive control or shared control of any addresses without his/her own consent.
\item $P$'s  secret keys have conjectural pseudomonopresence at all times.
\item $P$'s level of security is as high as the level of physical security that has been reached for the storage and
handling of $P$'s secret keys (which are only handled by the transaction constructor).
\end{enumerate}

$P$ has other worries than losing exclusive or shared control or conjectural pseudomonopresence. For a proper
functioning of the system $P$ needs to be sure that:
\begin{itemize}
\item The majority (in terms of processing power) of mining clients is working correctly according to the 
protocol specifications.
\item The user input client and the transaction client perform the required tasks (not blocked by a DDoS attack,
deadlocked due to a virus infection, or a similar problem). (The user input client and transaction placement 
client never handle any confidential information, and security issues for those components are  minor.)
\end{itemize}

Under these assumptions $P$ can be confident about secure operation of INCOs under his/her control.

This fact represents a remarkable strength of the Bitcoin design. By splitting the client in four components: miner, 
user input, transaction contractor, and transaction placement, and by using security by forward physical separation,
$P$ can work securely without the use of passwords, firewalls, security models etc. for any other purpose than
making sure that $P$'s user input client and transaction placement client are not blocked from operation.

In other words: a protocol for using an EXIM (or a TIM) has been sketched that
reduces security to a combination of three aspects: (i) sufficiently strong cryptography (mathematics), 
(ii) physical security (physics, biology, and psychology), (iii) correct computation of cryptographic functions on a 
sequential machine (computer science). 

\subsection{Substituting computer security for physical security}
The protocol for usage of an EXIM outlined in Section~\ref{Bcbss} is manifestly problematic from a 
practical point of view. Users prefer to be able to perform all processing on a single device.
Unless the single device is constructed as special purpose hardware that simulates multiple devices
there is no escape from computer security entering the story at a grand scale.

An additional objection against the protocol outlined above arises if one considers achieving 
physical security
a difficult task in comparison to achieving computer security. In some circumstances
the use of specialized hardware may be visible to other agents, and levels of monitoring may exits that
make physical security problematic. Then working on a single general purpose general computing device
may be simpler even if that requires the application of a highly sophisticated software based 
approach to computer security.

\subsubsection{Computer security enters the picture}
Because the protocol suggested in Section~\ref{Bcbss} is cumbersome and time consuming a user $P$ 
may want to perform all work on a single device. A user may plan to do all work on a single 
general purpose computer
which necessitates to shape by way of software the decompositions which were outlined in~\ref{assumptions}.
That ambition, which has nothing to do with informational money as such, 
brings into the picture an acute need for computer security.

Suppose that agent $P$ accepts that an INCO, say INCO$_{h}$ (hypothetical INCO),  for a particular EXIM, 
say EXIM$_{h}$ (for hypothetical EXIM) cannot be stolen by definition,
but it can be lost by the agent who thought to have exclusive control. Now we assume that 
EXIM$_{h}$ and Bitcoin (not an EXIM but a TIM) are contenders on the highly competitive 
market of informational monies.

Suppose at the same time that agent $P$ is a Bitcoin user as well.
$P$ may consider him/herself the owner of a particular quantity of BTC (Bitcoin INCO).\footnote{%
Within a TIM, being an owner of an INCO neither implies nor contradicts having control over 
that INCO. Similarly having access to an INCO (that is controlling that INCO) neither implies nor contradicts
ownership of that INCO. In other words, ownership and control are in principle 
orthogonal notions in this case. In practice one tries to make ownership and control
coincide by default.}  

Now relative to a TIM (or an MTIM) computer security can be simply expressed as the 
reasonable certainty that given sufficient physical security, the computer installation
used by $P$ is in such a shape that $P$'s INCO$_{h}$s will not be stolen by any other agent $Q$. The
puzzling issue is that the concept of an INCO being stolen from $P$ is almost independent of the notion
of computer security, whereas the notion of an INCO$_{h}$ getting out of control cannot be
understood in the absence of computer security. In other words: without satisfactory computer security in place
the INCO$_{h}$ is already out of control right from the moment of its creation inside $P$'s computer.

The problem we wish to put forward is this: what minimal awareness of computer security must
$P$ have as a user of EXIM$_{h}$? And is it indeed the case that as a user of Bitcoin, or of  any other TIM
that recognizes ownership of its INCOs, having that same
awareness is merely optional for $P$?

Another way of formulating the same issue: is it true that a Bitcoin user $P$ (assuming that 
Bitcoin is not an EXIM) can entertain a perception of his/her computer technology (including security)
at a higher level of abstraction than a EXIM$_{h}$ user because as a EXIM$_{h}$ user $P$ must have
a mental model of the mechanics of computer security? If that is true a Bitcoin user can do with 
information security (mainly consisting of cryptology, information theory, and probability theory), 
whereas an EXIM$_{h}$ user$P$  must in addition
think in terms of computer security (the dynamic security of real time protocols concurrently active 
in the multi-thread(s) running on $P$'s machine(s)). 
This amounts to the distinction between mathematics and computer science!

The distinction that underlies these questions may not appeal to all readers, but it emerges as
follows: the dependability of EXIM$_{h}$ rests on the correct implementation of a valid security model.
That a correct implementation exists, and is present in $P$'s devices is extremely hard to
judge and even harder to prove. For Bitcoin a valid security model need not go beyond warning
intruders that they are now working on foreign territory and that such is unwanted, and may be detected
and then may lead to adverse consequence for the intruders. In addition
entering $P$'s private sphere should not be made too easy. But the defense of $P$'s property
(INCOs owned by $P$) can be driven forward by legal means, prosecuting the intruder with 
means outside Bitcoin as well by means of steps available inside Bitcoin (for instance reversion 
of a transfer).

We conclude that the priority of access (control) over ownership in an EXIM can only materialize if
a user of an EXIM is  aware of a valid security model of which the implementation can be required
from a provider, who can be sued legally in case the security model was incorrectly implemented.

If an attacker finds a hole (weakness, flaw) in the security model that $P$ claims to use 
and that $P$ asked his computer system provider to comply with, then the attacker's resulting control
of an INCO$_{h}$ formerly under control of $P$ may be deplored by $P$ but it cannot be combatted 
through legal means. If the security model is valid (a hypothesis for which $P$ must be held responsible)
then two kinds of problems must be taken into account: 
\begin{enumerate}
\item $P$'s system incorrectly implements the security model (and $P$ may try to hold its manufacturer accountable),
\item the system is dependable but $P$'s use of the system is erroneous, 
in spite of the fact that $P$ has been correctly instructed on how
the system needs to be operated (and $P$ is to blame), or
\item the system is dependable but $P$'s use of the system is erroneous, 
which has been caused by the fact that  $P$ has not
been correctly instructed on how the system needs to be operated (and $P$ can try to hold the agency 
who delivered the equipment accountable).
\end{enumerate}

Summing up we conclude that EXIM$_{h}$ user $P$ needs to have a security model at hand from 
which it can be understood by $P$ that undesired interference by other agents (such as capturing
control of an INIT without $P$'s consent) is not realistically possible, whereas $P$ as an EXIM
user only needs to understand that an intruder must pay some effort for successful
intrusion and most importantly will be warned when trespassing into $P$'s digital private
``property''.

\subsubsection{Informaticology of computers and programs}
If one chooses to be dependent on a specific computing technology for money for years to come, we hold that it is
advisable to take as much a principled approach towards one's understanding of computers and software as it
is towards money and the corresponding social interactions. This viewpoint is not shared by many, at least not in the 
particular form that we are advocating. We hold that a systematic approach must be chosen to 
provide definitions of all core notions.
Such definitions may be simplistic when compared with existing technology, but at least they can serve as theoretical 
approximations.

For instance, if one arrives at the opinion that an OS is needed to handle a wallet on a device, one
should attempt to provide a definition of an OS. As mentioned above that is not so simple. Textbook definitions notoriously
say what an OS is for without saying what it is. If one needs computer programs that one can read, one needs a definition
of a computer program. Again not a simple matter if one discards nearly meaningless descriptions
like ``a set of instructions that does something useful''. The first author made an attempt to define what a program 
is in the project on program algebra starting with~\cite{BL02}. If one insists that multi-threading must be applied inside 
a P2P EXIM client one needs a definition of that. With a definition we do not mean an informal explanation of an existing
technology but a closed theoretical story that produces a definite concept, which might be applied as a first
approximation to more involved concepts needed to analyze existing or future technologies. An attempt to provide 
such  definitions for threads and multi-threads can be found in~\cite{BM07}.\footnote{%
Theories of multi-threading are hard to find for someone who disagrees with the seemingly dominant view that a 
theory of multi-threading is just a concurrency theory with the term process replaced by the term thread. 
The fact that most multi-threading takes place inside
deterministic machines has led to the notion of strategic interleaving as an alternative to arbitrary 
interleaving known from many concurrency theories.}
 
If one prefers a view of computer software as black box code, that is binary information which causes a machine to operate in
a certain way, without paying any attention as to how this causation might work, one arrives at the concept of control 
code that has been investigated in~\cite{BM2009}.

Obviously for each core notion in computer science alternatives exist to the approaches just mentioned. 
What is often missing in our  view is the objective  to provide definitions from scratch and to provide such definitions
in such a manner that a
notion or concept is constructed by means of that definition, and so that the rationale of a definition is 
not based on the assumption that
an audience already accepts the existence of a thing or concept or phenomenon and merely needs a reminder
of that psychological fact. Often, however, if concepts are built from scratch, that takes place in a very mathematical
setting where little attention is paid to the question whether or not the keywords used are usable as definitions
of the corresponding notions in computing. That approach is quite attractive because it leads to efficient
mathematical theory without the need to contemplate informal use of language. 

From the first author's experience with the development of formalized theory we wish to mention
the process algebra of~\cite{BK84} or the module algebra of~\cite{BHK90}. 
In both process algebra and module algebra
the question to what extent the notions of process and module as 
produced by those theories matches with common informal
understandings thereof in computing was not considered essential. 
Process algebra and module algebra constitute attempts to
develop applied mathematics for computing but not definitions of 
computational concepts per se. In that sense these projects
differ from the more recent development of program algebra in~\cite{BL02} and 
thread algebra in~\cite{BM07} where mathematical elegance
has been sacrificed on purpose to an attempt to design ``realistic'' concepts.  

\section{Bitguilder: transforming a TIM to an EXIM}
A TIM can be viewed as a piece of information technology only and be used under the guidelines that hold for an EXIM.
This way one transforms a TIM into an EXIM. We will write EXI(X) for the result of this transformation when applied to a TIM
X. If Y is a TINM then EXI(Y) produces an EXINM. Doing so for the specific case of 
Bitcoin  has the advantage that a hypothetical system emerges which
can be discussed more freely than the existing Bitcoin about which making some claims may be problematic for various reasons.

\subsection{Bitguilder = EXI(Bitcoin)}
With Bitguilder we will denote a hypothetical renamed copy of Bitcoin where BGUs (Bitguilder units, the equivalent of BTCs)
are stored and circulated instead of BTCs and such that the terminology (ownership making way for control) 
has been adapted to that of an EXIM. 

Bitguilder is positioned as an EXIM, because we view Bitcoin as a TIM.  If Bitcoin is 
considered a TINM, Bitguilder is correspondingly cast as an EXINM.

\subsection{On the ``intended rights'' of a Bitguilder user in case of difficulties}
We cannot say that a Bitguilder user has rights unless those have been shown to exist  and persist in time. But we can
summarize intended rights as an additional specification of an Bitguilder system, from the user perspective.

Below $P, Q, Q_1$ and $Q_2$ are users  of Bitguilder clients; $A$ is an attacker of the system 
who might misuse or disregard $P$'s interests;
$M$ is the security model that $P$ has in mind for $P$'s system; $M$ is supposed to be satisified by 
implementations of a specification $S$ (for Bitguilder clients)  that was delivered by an open standards and community driven
specification process; software provider $E_{s}$ has engineered the software client that $P$ is using 
(from a specification $S$) on hardware and middleware that
has been engineered and provided by $E_{h,m}$. Finally $B$ is an external observer.
\begin{enumerate}
\item If $A$ intrudes $P$'s device and by doing so obtains\footnote{%
We will also say that $A$ captures control of the BGUs. However, it is not easy to define when $A$ starts 
having access to the BGUs that $P$  expected being in sole control of; 
once $A$ understands how $P$ can be attacked, or once $A$ plans to perform the attack, or 
only after the planned attack has been successfully performed by $A$.}  access to $q$ BGU which are 
moved out of control of $P$ then $P$ may
hold different parties accountable:
\begin{itemize}
\item $P$ itself was to blame  if the security model $M$ that $P$ claimed to use has a flaw that has been exploited by $A$,
\item $P$ was at fault if $S$ does not guarantee that $M$ is satisfied (the open source process gives no guarantees),
\item $E_{s}$ if the flaw exploited by $A$ lies in the implementation and if $E_{s}$ has signed for an 
appropriate product responsibility, (otherwise $P$ is to blame),
\item $E_{h,m}$ if the computer system was not working correctly, thereby enabling the attack,
\item $A$ for intruding in $P$'s private informational sphere. 
\end{itemize}
Under no circumstance, however, the penalty for the agent claimed guilty depends on the numerical value of $q$.
Moreover $A$ is entitled to subsequent use of the amount, simply because $P$ has captured control.
\item If $P$ performs a double-spending attack on $Q_1$ and $Q_2$ then $P$ can be held 
accountable and be punished or fined.
If either one of $Q_1$ or $Q_2$ receives the amount that user will not be 
required or forced to send it back by legal means
(of course it can be asked, and it may be done if that suits the relationship management of the receiver in case).
\item If $P$ uses $Q$ as a participation service provider, where $Q$ serves as a Bitguilder client instead of $P$, 
then $P$ may claim to own BGUs, though
in fact $P$ only has a claim on $Q$ that can be expressed in terms of BGVUs in the 
terminology of Paragraph~\ref{BGnotation} below.

\item In the case that $Q$ acts as a participation service provider for $P$: if $A$ intrudes $Q$'s system and gets 
access to $q$ BGUs that $Q$ is in control of as a service to $P$, so that
$Q$ loses control of these BGUs in favor of $A$, then by default $P$ may claim a restitution from
$Q$ that matches the full amount of $q$ BGU determined at the time of the occurrence of the capture by $A$.\footnote{%
This clause indicates that it is important to define at what time exactly an amount is 
being captured by an intruder, or by a prospective intruder.} 
Arrangements may be made that determine another
distribution between $P$ and $Q$ of the risk of loss of control by $Q$.
\item $P$ may outsource the handling of its BGUs to $Q$. In that case $P$ shares access to 
these BGUs with $Q$, and for that 
reason $P$ cannot hold $Q$ accountable if $P$ loses control by either mistake of dishonesty of $Q$. 
(BGU access sharing implies trust!)
\item $P$ can always prove to $B$ having control over a specific amount through an underlying account, but 
$P$ has no method for demonstrating to an external observer $B$ that $P$ is not in control of a specific account. 
In particular $P$ cannot prove exclusive access, and cannot exclude shared access with any other use $Q$.
\item If $P$ is in control of an amount $a$ and proves that to $B$ it is not up to $B$ to classify $P$'s 
access as either black (illegal) or white (legal).

\end{enumerate}

\subsection{Options for additional Bitguilder features}
Having available Bitguilder as a hypothetical EXIM the following question is enabled: what more features 
could be usefully added without adversely affecting its status as an EXIM? Here are some options for extending Bitguilder
to an even more hypothetical system BitguilderPlus:\footnote{In no way we intend to suggest that Bitcoin 
would profit from any of these extensions, because that assertion requires us to pretend without justification
an awareness of Bitcoin design objectives. Clearly once a certain Bitcoin ideology has been chosen, some of these
suggested extensions may qualify as potential improvements.}
\begin{description}
\item{\em FemtoBGUs.} Because there may be a fundamental need for Bitguilder 
micropayments it seems useful to have much
smaller units than Bitcoin provides: 
the FemtoBGU ($10^{-15}$ BGU) might be used as a minimum of transfer and as a minimal amount that an 
reside on an account.
\item{\em Future transactions.} A transaction may be performed in such a way that 
the amount is removed from the sender's
public key but arrives at the receiver's key at a later stage only. This allows a receiver to be sure that the transaction
will take place while it guarantees the sender that the receiver cannot use the amount before a certain moment. 
The transaction indicates to all Bitguilder participants that the receiver has confidence in the stability of the system.
\item{\em Conditional future transactions.} Future transactions may be verified even if amounts are insufficient 
under the condition that such transactions will be effectuated as soon as amounts connected to input accounts suffice.
\item{\em Intentional key pair destruction.} If $P$ is in control of a key pair for which there is no use 
anymore (no incoming future transactions have been verified for it) $p$ can simplify his/her task of 
storing private keys by
officially destroying (perhaps discarding is a better term)  the key pair with the effect that 
never in the future a transaction into that public key can be made.
\item{\em Customized public announcement of system user policies.} Bitguilder users may experience 
unclear conditions asking for clear policy guidelines, 
and such guidelines may need customization. 

Here are some examples of policies that might be announced (not all combinations make sense, however), 
viewed from the perspective of user $P$:
	\begin{description}
	\item{\em Restitution policy.} If $P$ receives an amount from an unknown (for $P$, perhaps 
	not for the tax office or the police) agent $X$ (and with unknown purpose for $P$), then 
	there must be a fixed method available to $P$ for sending the amount back so that no 
	fees are sure and no costs arise for $P$. 
	
	This suggests that:
	\begin{itemize}
	\item  a transfer of a single 
	Satoshi is problematic, because a corresponding fee has at least the same size, and transactions without fees will
	make the system vulnerable to attacks from dishonest users,
	\item restitutions must allow smaller 
	transfers than ``ordinary transfers'', and
	\item there are no restitutions of restitutions.
	\end{itemize}
	
	Justification: if $P$ receives an amount from a (possibly) criminal organization $X$ revealed 
	by the use of a public key only then $P$ cannot prove not to know anything 
	about $X$, so $P$ needs a way to demonstrate not profiting from $X$ and restitution seems to be the only option.
	\item{\em Open identities and objectives.} $P$ can opt to state on a publicly visible website some 
	or all of the following policy elements (technically to be understood as promises in the sense of 
	\cite{Burgess2005,Burgess2007,BergstraB2008}):
		\begin{itemize}
		\item that $P$ accepts anonymous transfers into a specified account (or set of accounts termed gift accounts)
		which are supposed to be gifts 
		meant to support $P$ in achieving a set of publicly stated objectives,
		\item that $P$ will publish for 
		every outgoing transaction the name of the recipient and some information about the purpose of the payment, 
		\item that $P$ reserves the right for 
		every outgoing transaction to publish 
		the name of the recipient and some information about the purpose of the payment, 
	 
		\item that $P$ will publish (or merely reserves the right to publish) the name of each agent that 
		sends it an amount (excluding gifts) and some 
		information about the purpose of the transfer (and lacking such information the amount will be restituted),
		\item that $P$ will publish information about its restitutions (not needed if such facts can be obtained from the block
		chain after reengineering of its technology),
		\item that $P$ will publish all accounts that it claims to have control of,
		\item that $P$ is willing to prove for each of its public keys that it is in control of the corresponding secret key
		against a fee (a restitution mechanism will automate that feature),
		\item that $P$ will publish for each of its accounts (public keys) if it has lost control 
		(either by losing the secret key or
		by finding out that another agent now possesses the corresponding public key as well),
		\item publishing comprehensive information about shared control of the keys it controls,
		\item a policy to ensure that certain transfers and policy changes are performed in case of $P$'s death.
		\end{itemize}
	\end{description}
\item{\em Transferral of (partial) wallets.} Complex delegation of tasks may require the need to send 
another agent one's wallet or part of it, that is to communicate secret key information. 
Such actions may be included in the top level design of the system, 
rather than being engineered alongside it.

\item{\em Anonymous transactions.} Bitguilder is as weak as Bitcoin concerning anonymity. 
Obviously more support for anonymous transactions can be engineered on top of 
Bitguilder when initialized as a technical clone of Bitcoin.\footnote{%
We don't want to state by any means that such engineering of increased 
anonymity support would improve Bitcoin. 
Additional support for anonymous transactions will make Bitcoin more attractive to some users, 
while at the same time such additional support may be perceived by other users as a highly problematic 
feature that introduces unmanageable risks.}

It is not easy to explain in simple words what technical requirement on a system is meant if it is to provide users
the opportunity for anonymous transfer. All theory of anonymity is based on Chaum's notion of an anonymity set and 
defining that set convincingly is already non-obvious in the case of Bitcoin (or Bitguilder).
In \cite{MiersGGR2013} Zerocoin is described as a P2P informational 
money which gives users an opportunity for anonymous transfers. 
Similar adaptations can be applied to Bitguilder of course.

\item{\em Parallel units within the same system.} It may be useful to have a second unit with the same 
system that operates in parallel.\footnote{%
Having different monies in parallel is current practice, but the feature meant here pertains to a functional
basis for distinction between monies instead of a regional basis for their distinction.} For instance:
	\begin{description} 
	\item{\em Quantity management.} An additional unit, say QMU (for quantitatively managed unit)
	plus corresponding circulation system and an exchange market
	to and from BGUs of which the value fluctuations are somehow moderated. 
	The total circulation of QMU may be managed
	in a flexible according some assessment of demand for QMUs.
	
	\item{\em Dedicated units.} It may be useful to support a portfolio of dedicated units, 
	e.g. linked to emission rights or other resources for which certain restrictions to free trade 
	must be imported. This opens the way for providing  amounts (expressed in dedicated units) for 
	exclusive use in say healthcare, or education. Local units may be facilitated as well.

	\item{\em Ownership based units.} If additional units are provided some may be 
	outside the EXIM regime and within the TIM regime
	thereby allowing  users of such units to rely on legal protection of ownership 
	instead of access control and technology.
	\end{description}

\item{\em SecureBitcoin basis.} Security in BitguilderPlus can be improved 
(compared with the Bitcoin-like security of Bitguilder)
 by using digital signatures with higher entropy and by
making use of more involved hashing for proof of work calculations. 
Such strengthened technique can be imported in Bitguilder.

\item{\em Mining democracy.} The unpredictable future behavior of the mining community constitutes both a 
strength (or at least an intriguing experiment) and a weakness (risk) of Bitcoin.  
Here are some thoughts on how mining might be adapted. None of these modifications are obvious and all may 
have disadvantages that potentially outweigh any potential advantage.

\begin{description}
\item{\em Programming competition instead of processing competition.} Instead of computing very fast, 
which slowly but steadily dismantles the original level playing field for miners, 
the competition based on proof of work might be organized on the basis of a virtual machine 
model, where a program is executed
and its execution time is measured by means of a formal model. Then the competition is for 
programming competence instead of computer investment. This kind of competition might include novel machine architectures given
agreed simulations of those.
\item{\em Stronger mining curve sloping.} It may be useful to have mining difficulty in BitguilderPlus grow even faster 
with total block length than in Bitcoin 
in order to discourage attacks on the block chain encoding of agreed history more effectively.

\item{\em Penalties for double-spending attempts.} Once a double-spending attack has been spotted that fact itself 
can be represented in the block chain and the account from which that particular attack place can be blocked forever. 
The penalty thus collected may thereafter be allocated for further mining.\footnote{See~\cite{EveraereST2011} 
for technical details on fining  strategies.}

\item{\em Early user advantage moderation.} The formidable advantage of early Bitcoin users
constitutes a serious difficulty for which we cannot easily suggest an elegant solution. 

The role of mining in Bitcoin creation and initial Bitcoin distribution is possibly too prominent. 
Perhaps it is possible to postpone the phase where early mining gives quick and cheap results. 
Mining may also be made more function, for instance blocks containing verified double-- (or triple-- etc.)  
spending attacks, thereby publicly blaming the addresses from which such attacks were launched, may 
win out against blocks failing to make and verify such observations.

It is a remarkable property of Bitcoin that early users with low tech equipment had a real chance to be successful with
mining, simply because everybody understood Bitcoin clients as monolithic software components to be 
run on general purpose equipment available to the general public. Only after some time mining became specialized
and inaccessible for less sophisticated users.

A reasonable step forward might work as follows. Assuming that BitguilderPlus has acquired very strong institutional 
support then it may be meaningful
to prescribe an initialization phase under centralized control where each citizen is provided with a key pair and some 
initial volume of BGUs, that is a coin. Thereafter every newborn citizen is issued a similar loaded key pair 
by means of an automated transaction that uses so many biometric data as are needed to determine unambiguously the
birth of a new citizen. A certain percentage of the transaction fees is continuously withdrawn as a taxation in order to
pay for these new allotments. The total amount of EXICs can be made dependent on the number of persons alive, 
though perhaps not linearly so in order to work against overpopulation. 
\end{description}

\item{\em Agent identity management and authentication.} It is a remarkable feature of Bitcoin that the 
classical notion of  trust plays no role anymore. That may be seen as too extreme and for might be considered a weakness
of the Bitguilder architecture. One may imagine besides or as a component of an 
EXIM a system where agent identities are maintained and verified. On a voluntary 
basis an agent may wish to store biometric data that allow
full identification and authentication. Let us assume that such a system can be developed on 
principles not too distant from the principle on which Bitcoin is founded.

\begin{description}
\item{\em Rewarded competition for each Bitguilder related information processing task.}  Once a collection of verified agent 
identities is emerging the task arises to link public keys to such identities. Of course agents may reveal the public keys
they are  controlling of, and they may use the system is used for the validation of such claims, 
but many agents may not wish to provide complete information. 
This creates a task for external agencies eager to know who did what. 
Instead of leaving this task to external agencies it might be considered an important 
processing task for which BitguilderPlus provides first class support. 

With such mechanisms in place Bitguilder can be helpful for anonymous operation. 
The reasoning then works as follows: as long as no participant has claimed 
(and proven including peer verification by way of mining) a link between an 
agent and a public key controlled by that agent, that link may be  considered (by that agent) unknown  to external observers. Of
course the latter inference depends on the agent's trust that this part of the 
BitguilderPlus related IT competition has state of the art participants.

\item{\em Trusted agents.} Trusted agents may be enabled to perform transactions and have the services to be 
acquired in return provided to them before these transactions have survived the mining based verification process. 

\item{\em Guarantees against double-spending by trusted agents.} An agent $P$ may arrange (in a verified way) that
another trusted agent $Q$ guarantees $P$'s transactions. That is, if an agent 
$R$ (who trusts $Q$) upon receiving a transaction from $P$ turns 
out to be a victim of a double-spending attack attempted by $P$, then $R$ may claim the damages from $Q$, instead
of from $P$. The system must ensure that $Q$ is always able to back the transactions made 
by each $P$ who made this arrangement with $Q$. The advantage of this arrangement 
may be that $R$, knowing of $P$'s backup by $Q$ may deliver
a service before having observed a certain threshold level of blockchain stabilization that would be required in the 
absence of a guarantee.\footnote{%
New mechanisms for preventing the latency of Bitcoin transaction validation, in order to facilitate fast payments are
already  an important topic, we mention the marker addresses (formerly called 
green addresses) of~\cite{Vornberger2012} as an example.}

\item{\em Licensed mining.} The design of Bitcoin allows no role at all for non-automated decision taking. Now it
is conceivable that this level of distrust cannot support a satisfactory  mining industry in the long run. 
If instead of having many anonymous miners there might be a P2P based voting system licensing approved miners.
Some 500 licensed miners representing many interests worldwide might create and serve a very stable system.\footnote{%
Licensing is very much
part of the open source software ideology, but licensing on the basis of demonstrated QoS is not. If delivering 
proven QoS can be checked automatically, the open software proponents may come to like it nevertheless.}

\end{description}

\end{description}

\section{Natural kinds and speculative Bitcoin impact analysis}
In this Section we will provide an assessment of the extent to which Bitcoin might influence the development
or evolution of informational monies. This question
is interesting especially if Bitcoin will give way to some other informational money. When that happens the impact of Bitcoin can
only be assessed indirectly. Clearly the question cannot be reliably 
answered to date but we hope to demonstrate below that some provisional progress can be made. 

In order to approach this issue some theoretical approach must
be borrowed from elsewhere. We will make an attempt to apply the philosophical ``toolkit'' of natural kinds that has already
been applied with some success to the theory of biological evolution. We will follow the style of that application of natural kinds, 
acknowledging at the same time that our discussion may be felt to be not quite in touch with recent philosophical insights
about the status and role of natural kinds.

Some perspective  on ``what Bitcoin is'' must be chosen beforehand. 
We will consider its essence to be a computer program, or rather a collection of computer programs each developed from
preceding members of the collection and ultimately all descending from Nakamoto's original software creation. We will
understand computer programs in a dual way, first of all as sequences of instructions.\footnote{%
Programs as instruction sequences is a view that might underly an interpretation of
``computer program'' as a natural kind, which we do not subscribe to at this stage.}
A second perspective of programs is
that they consist of groupings of algorithms each of which can be understood both in functional terms, and also
as computer programs often best looked at from a  higher level of abstraction.  

Software is often considered in evolutionary terms (e.g. see \cite{SinghS2013}). Through ongoing revisions and improvements computer
programs evolve under competitive pressure. An open source programmer gives any new version of a program program 
the capability of having offspring by inviting  all open source programmers  to its survival by creating subsequent
descendants, hopefully with improved qualities, or with better adaptation to a changing environment. 
Comparing a line of software with a biological species is reasonable
to the extent that we will propose to have the link between software and natural kinds arranged in more or less 
the same way as it has
been arranged in the case of biological species.

A marked difference with a biological species is that individuals of a software species are programs of which multiple
copies can exist simultaneously. We will consider all copies of a program, each copy usually resulting from downloading the 
program from a ``master copy'' on a repository that is hosted on an appropriate website, as being ``the same'' program, with the 
particular species of programs consisting of an entire chain of subsequent versions derived from an original program. In
biology genetically identical twins do not result from copying but from low probability commonalities between two 
production processes. At this point the comparison between programs and individuals 
of a species becomes weaker: programs will carry on forever unless new versions are needed because changes must be made. An aging based life-cycle necessitating
a reproduction process for species maintenance is not present during the life of a software species.

\subsection{Preparatory remarks on natural kinds}
The notion of a natural kind has been analyzed in depth in philosophy for several centuries. We will briefly attempt to
state what a natural kind is, acknowledging the wide variation in philosophical views about natural kinds.\footnote{%
The sketch of natural kinds that we provide embodies essentialism of natural kinds, which became popular though the 
influential work of Hilary Putnam (\cite{Putnam1975a,Putnam1975b}, see also \cite{Dupre1981}, \cite{Boyd1991},
and \cite{Reid2002}.} 

A natural kind $K$ is a kind (that is some class of entities) with the following additional characteristics:
\begin{description}
\item[\em Objective existence.] $K$ has instances and every instance $k$ of $K$ has an intrinsic 
existence that can be conceived without the existence 
or the appreciation of human beings. (It is not required that the grouping together of these $k$'s into $K$ is 
independent of human action or thought.)
\item[\em Essential and defining micro-structure.] Being a $K$ is to be thought of in terms of the 
essential structure of what something is. The microstructure is defining for the natural kind ``by definition'', even if it
is not entirely known and to some extent still a matter of speculation.
(For instance 375 is the sum of 375 units 1, and a bottle of water deserves that description 
because of the particular molecular structure of its content. Gold was available as a natural kind 
before the physics of its atoms was known and  its atomic essence was conjectural only).
\item[\em Division of work.] Those who apply the notion of $K$ in order to speak of individual $k$'s and their use need not have 
full and certain knowledge of what it (essentially) means to be a $K$. Users of $K$ may differ from scholars of $K$ who try 
to get deeper understanding of $K$'s.
\item[\em Subject of scientific assertion.] $K$ is a concept about which research findings can be (and have been) 
stated.\footnote{%
There are millions of $K$'s is a possible finding. There are no $K$'s is a problematic ``finding'' because 
in that case $K$ does not qualify as a natural kind to begin with but merely as a nominal kind.}
\item[\em Topic of scientific investigation and discovery.] Research on $K$ may always introduce new 
insights into its structure and essence. Research on the essence of a particular natural kind will 
never be completed, new findings about it are always conceivable. 
Use of $K$ in reasoning and of $k$'s in practice is done at a level of abstraction that
may survive discoveries about the essence of $K$ and its instances. 
\end{description}

Many modifications of this notion of a natural kind have been contemplated, with blocks of gold remaining a prototypical
example that seems to survive all research on natural kinds. The question if natural kinds are also a natural kind has been raised
an the answer seems to be negative on that matter. The definition just given needs being improved for instance because 
many natural kinds seem to occur in clusters rather than stand alone.\footnote{%
If one experiments with the species of a cat as a candidate natural kind one soon finds that it is more plausible to assume that
male cats and female cats constitute a cluster of two natural kinds. However, the view that biological species constitute natural kinds,
has been replaced by a rather more sophisticated alternative view based on genetics and traits with potentially
longer life-lines than individual species.}

Probably it is an illusion to view
natural kinds and surrounding ideas as a tool from the ``applied philosophy toolkit'' that may be
applied  by individuals from outside professional philosophy in the hope of 
clarifying their descriptions of novel phenomena. 

Having said that, we notice that
there seem to be many occasions where a ``natural'' and informative use of the notion of a natural kind can be 
made which only suffers from the weakness that some recent paper in philosophy can be found undermining just that
picture of what role a natural kind really is supposed to play in a scholarly account. 

\subsection{Natural kinds in the context of Bitcoin}
 By viewing Bitcoin as a species of programs for implementing informational monies, its evolutionary 
significance can be evaluated only in the light of subsequent
related programs. Thus, rather than analyzing its potential impact on the development of informational monies, 
we will limit our discussion to the impact that Bitcoin may have on subsequent (software) 
implementations of informational monies. 

At some stage biological species were
understood as prime examples of natural kinds, but that point of view has now been superseded. 
From this course of (philosophical) events we conclude that
it is not plausible to consider a specific money (be it Euro or Bitcoin or something else) as a natural kind.\footnote{%
One might object that if classifying Bitcoin as a natural kind, say of informational coins, is problematic, its 
classification as an element of a natural kind, say of TIMs, must be contemplated. We claim that the same objections 
apply that have been raised in
biology against the natural kind status of an individual biological species:  
lack of clear demarcations between different species because of gradual evolution from species to species. 
Recently artifact kinds have been studied in order to deal 
with conditions where kinds of artifacts can be profitably compared with natural kinds. We don't believe that an analysis
of Bitcoin in terms of artifact kinds is necessarily more convincing, mainly because of its mathematical background.
Obviously we cannot exclude that analyzing the impact
of Bitcoin through artifact kinds is an option that may prove more rewarding on the long run than approaching the same issue
with natural kinds.}

Rather than looking at individual biological species one may consider so-called traits, that is features and capabilities, 
often based in appropriate combinations of genes, as the prime examples of natural kinds in biology. 
Traits may start to exist, that is a first instance may come into existence, at some stage, perhaps due to some lucky
genetic mutation, and may subsequently cease to exist through natural selection. 
In this conception the natural kind (trait) has as its instances
combinations of genes that are physically present within an individual of a species and which cause certain features or capabilities
of that individual to pertain.

Externalism applies and division of labour applies now as follows: 
some may use natural kind terms to indicate traits (as well as their particular implementations as embedded
in sections of the genetic base), while others (in the division of intellectual labor, 
using the natural kind terms as an inspirational source) 
are investigating these mechanisms in order to come up with better accounts of the essential content of the 
traits at hand. More specifically, the essence of a particular natural kind (i.e. trait), perhaps understood by its ``users''
in terms of some capability that instances of the trait cause to exist in an individual of the species, can in principle 
be discovered to be different from existing assumptions, 
for instance because an additional gene has been found to be necessary for the mentioned
capability to be created in an individual.

Specific natural kinds thus conceived can survive the transition to another species and the splitting of 
species in different successor species. These natural kinds
may have an extension consisting  of two different types of instances:
\begin{itemize}
\item  instances of different genetic mechanisms that have 
been developed along
independent evolutionary paths for solving the same operational problem (e.g. flying, or vision in 
some part of the spectrum,
or vision at some distance, or measuring the speed of motion of another agent), and 
\item instances of gene combinations causing slightly varying operational capabilities that 
have evolved from the same original implementation. 
\end{itemize}

\noindent These ideas can be translated to the evolution of programs. 
Informational money related natural kinds (IMRNKs) are features or combinations of features that may (i) originate at some stage,
(ii) become extinct at some stage, (iii) be handed over from money to money during a number of evolutionary
phases. Instances of such natural kinds are combinations of fragments of programs that together embody the algorithms
responsible for realizing these features or combinations opt features.

The role of Bitcoin (or any money) in the evolution of informational monies can be approached via the following questions:
\begin{enumerate}
\item Which IMRNKs are candidates of being originally introduced in Bitcoin? 
\item Which, if any, IMRNKs are re-introduced in Bitcoin, after having diapered from existing software product lineages. 
\item Which IMRNKs are candidates of being qualified as extinct because of their 
absence  from (or reduced role in) Bitcoin?
\item Which IMRNKs are made (or seem to be made) more prominent by their role in Bitcoin?
\item Which IMRNKs are made (or seem to be made) less prominent by their role in Bitcoin?
\end{enumerate}

Here are some possible answers to these questions:\footnote{As far as we can see these answers are 
consistent with the conclusions formulated in~\cite{Barber2012}.}
\begin{itemize}
\item Bitcoin signifies a definitive endpoint to the IMRNK that requires that money-items have 
some real and intrinsic value (other than 
the value as money-items), or that money-items are descendants of items that have such non-monetary value.
\item Bitcoin creates a world in which the full history of each quantity of money is entirely 
known and transparent, modulo the IP numbers of machines on which key pairs were 
created and digital signatures were computed. Instead of ``what happened with this 
banknote'', one will now ask ``who effectuated those transactions'' or who is in control 
of that account?" Bitcoin always maintains a very well-defined abstraction of the past 
and current status. This constitutes an IMRNK that will not go away, only the level of abstraction may differ for future monies.

Following \cite{Bergstra2013a} this IMRNK can be phrased alternatively as follows: Bitcoin captures the concept of a money
inclusive a fully worked out theory of circulation, as well as a compelling proposal of what abstraction of the circulation history
needs to be taken into account. IP numbers of machines from which transactions were made are forgotten, and so a user 
names working from these machines, and motives they had for transferring, and so on. All that is not part of the concept of 
money in the world of Bitcoin.
\item Bitcoin brings mainstream minting (an IMRNK that seemed extinct) back to the public. 

Bitcoin terminology speaks of mining instead of minting
and seems to view minting as a process involving a central authority that Bitcoin seeks to avoid. In addition mining has the 
connotation of an upper bound to what can be found which minting has not. The argument can be reversed in that 
minting follows after mining (both in classical metallic monies and in Bitcoin).

\item Bitcoin introduces (potentially unlimited) divisibility of its unit plus potentially unbounded deflation as the exclusive 
mechanism
to deal with an increased need for money-items. A highly speculative and potentially controversial, but certainly original 
IMRNK. This option has always existed in principle but for monies with physical money-items of fixed unit size and above,
 the internal contradiction between physical wear (moving coin value down)  as  a consequence of  
use and potentially unlimited deflation (moving coin value up) is hardy practical.\footnote{%
If arbitrarily divisible Bitcoins are used as a money of account, one may profit from a notation 
for rational numbers in which
division is a total function having its own operator symbol besides zero, one, addition, subtraction, and
multiplication. Such a notation is found by taking BTC quantities as positive elements of the signed cancellation 
meadow of rational numbers
(see~\cite{BBP2013}).}
\item Bitcoin portrays a money as a dedicated P2P network based on a general computing infrastructure
which can be implemented by open source software that is maintained by an open 
source software oriented community. A 
novel IMRNK (or rather a so-called cluster natural kind) that seems to have no counterpart in previous monies. 

Included in
this IMRNK is that Bitcoin promotes the mutual trust of the user community (both in that communities' 
current operation, 
and in its future development) to the forefront as the exclusive basis of its justification 
as a locus of money.\footnote{%
A striking consequence of this IMRNK is that, although written specifications are still missing, everyone
can acquire an understanding of how the Bitcoin system works. This is in sharp contrast with conventional finance
where the idea that an average citizen knows the mechanics of the system, such as the 
special role of banks, fractional reserve banking, the interplay between banks and a 
national central bank, or the interplay between the national central bank and 
regional as well as worldwide supranational institutions, has been given up entirely. 
The price paid for this transparency is that a Bitcoin user must take notice of the 
development of software technology (Bitcoin clients) and 
supporting computer technology. That is a daunting task.}
\item Bitcoin introduces a commitment to a specific combination of cryptographic technique. These methods are 
constitutive of Bitcoin as it stands and not mere features of its current implementation (as it would have been the case
with all informational versions of preexisting monies). This bundle of commitments qualifies very much as a MNRC (or its
cluster version) because it is so obvious that it may become extinct through natural selection in some not so distant future
(quantum computing may introduce the need for an alternative to ECDSA elliptic curve cryptography and SHA-256 may
become superseded as the problem family of choice for the proof of work feature).\footnote{%
Although one may view ``being a cryptocurrency'' as the abstraction that lies behind this IMRNK, 
it is the appearance of
specific cryptography in the monies' specification that seems to be of paramount importance. 
That is not an implementation 
detail in any way. It is like vision in biology: although a species may acquire the IMRNK vision, that IMRNK is 
in its actual occurrences not ``abstract'' but it is
genetically committed to a part of the optical spectrum, and to a part of the intensity spectrum, and perhaps also
to geometric constraints. That commitment is intrinsic to the IMRNK.}
\item Bitcoin detaches money from geography. Not a new idea, but in technical terms Bitcoin 
may be considered an 
original source for this most important IMRNK.
\item Bitcoin discards some  anonymity features while strengthening other anonymity features (the overall 
picture is hard to assess).
\item Bitcoin strengthens irreversibility of transactions and the autonomy of users.
\end{itemize}

This listing is preliminary and incomplete. Nevertheless, our  impression is that even if as a 
combination of IMRNKs (that is as a species of implementations of informational 
money) Bitcoin may perhaps not be so attractive, its evolutionary importance, in terms of providing
an update of the portfolio of IMRNKs considered essential to programs for the realization of informational monies, 
may become very significant.

\subsection{Expected impact of the Bitcoin P2P network}
The decomposition of Bitcoin into a family of natural kinds some of which may outlive Bitcoin as a species of open source
technology (coupled with  a range of closed source relatives) provides a mechanism that can project on Bitcoin an influence
on the development of monies that outlives its impact as a software technology species. Speculation about the future of Bitcoin 
is a very popular matter on the internet these days. It seems that a single BTC must rise to a value of at least 50,000 Euro
for Bitcoin  to become
an influential mechanism on the financial world market. So either Bitcoin disappears from the scene or the BTC 
acquires an astronomic value. This strange contrast produces an irresistible attraction. 

Predicting the future of Bitcoin as a system, rather than of the natural kinds of which it is made up is probably harder. 
An outsider tries in vain to assign some underlying value to a BTC and must conclude that this is simply impossible. 
A value of 100,000 
Euro for a BTC is not intrinsically implausible, it is only difficult to believe that the BTC will ever make this 
much progression. However, as soon
as enough people believe that that wil happen it wil indeed happen.

Whether or not an economy with Bitcoin as the only system for the circulation of money can assign to a BTC a stable value
is unknown. It is a probably matter of experimental economic (and psychological) research  to find out under which circumstances 
the BTC can have a stable value. That kind of research can be extended to so-called dual 
systems where Bitcoin is not the only (near-)money around. A financial system where Bitcoin plays a role in a
bundle of monies each with different objectives seems to be an attractive perspective. Below we will consider a dual system. 

Following Gesell  an extension of the dual system with local monies with artificially 
enhanced inflation to prevent users from hoarding may complement the picture. 
Regional monies with built in deprecation, or negative interest (the way in which Gesell produces
inflation) can be shaped in an information style just as non-local monies can be.  
One may imagine a hierarchical system where
lower near-monies have more inflation on purpose. The economic question is: can such architectures constitute stable 
financial systems. At this stage there seems not to be the beginning of an answer to this important question.

\section{Interest prohibition: a dual system proposal}
\label{IntProh}
In this Section we will speculate on the impact that Bitcoin might have on the future development
of less conventional monies, in particular on the development and architecture of monetary systems in which 
interest payment is to be avoided, with modern Islamic finance as a prominent example. It seems that a significant 
connection between EXIMs and avoiding interest can be found.

In very classical views of money, some of which still persist,  payment of interest is seen with suspicion and debts are 
preferably avoided for that very reason. In~\cite{BM2011}  a money without interest is understood as a
so-called reduced product set finance (RPSF). In an RPSF some financial products are missing, compared to conventional
systems of money. In~\cite{BM2012} there reader may find further reflections on the possible merits of having 
monies without interest. A connection 
with Bitguilder (with unit BGU) may be as follows: Bitguilder is another RPSF (be it a hypothetical one) 
by its deletion of ownership in favor of access and control.

Another connection between Bitguilder and absence  of debts and interest may be found as follows. Following the 
line of thought of the original
motivation for Bitcoin as expressed in in~\cite{Nakamoto2008} one may hold that 
Bitguilder produces money whereas Euro's and other 
conventional monies do not, for the reason that such other monies pay lip service only to the store of 
value feature of money. Lacking the backing from gold or a similar good of independent value, the Euro and
similar monies fails to meet the requirement that a money provides a storage of value. Dropping the gold standard
can be seen as an step that had to be matched with another way of guaranteeing the nonoccurrence of catastrophic inflation. 
Modern monies have failed to develop just that. From that point of view an EXIM,
say the hypothetical Bitguilder, provides ``real money'' (in the sense of its definition, rather than in the 
sense of what is used on the street) whereas the Euro provides a barter catalyst only. Barter Catalysts like the Euro can
be modernized into the form of informational
near-monies, which may either be shaped as technically informational near-monies TINM in the terminology of 
Section~\ref{TINMdef}) or as exclusively informational (EXINM), or far more plausibly as managed TINMs (MTINM). 

As a near-money besides an EXIM an MTINM can be used to solve the social problems that governing bodies face. In an
MTINM some form of non-automated governance may be used to find a balance between inflation and deflation 
of the unit of the underlying TINM. An MTINM may satisfy the additional constraints 
mentioned in Paragraph~\ref{Wim} which are needed in order to enable societal governance to 
satisfy the needs of the population. 

\subsection{Dual system: Bitguilder + NMcoin}
Having made these design decisions a simple picture emerges. In this picture Bitguilder and one or more 
(M)TINMs can (and even should) coexist. We will consider the case that besides Bitguilder there is a second system, say
NMcoin, an MTINM with unit NMC.\footnote{%
Obviously if the objective were to propose or even develop a new architecture for Islamic finance then
Bitguilder will not be a plausible name for the EXIM to be incorporated in the dual system 
resulting from that endeavor. However, a dedicated abstraction
of Bitcoin towards an EXIM (say Bitguilder)  followed by a redesign (and an appropriate renaming) of that EXIM that takes into 
account (i) additional requirements imposed by  a specific philosophy of money, (ii) its role in a certain kind of 
economic system under the assumption (iii) of the presence of (the successor of) a previously existing 
money now cast in the role of a near-money rather than in the role of a money, seems to be feasible.

These considerations suggest a conceivable course of action for the development of Islamic banking: 
(i) move into Bitcoin to see how it works, as well as to participate in its further development, 
(ii) redesign it (say into Cresent-Star-Bitmoney) so as to accommodate important 
principles of justice such as a fair beginning and to reintroduce some dependence on trust, 
(iii) get Icoin started before Bitcoin's position (or the position of any Bitcoin competitor yet to emerge) 
has become too strong to admit the survival of other  related initiatives, 
(iv) formalize the principles of a dual (near-)money system.}
In this dual system it is to be expected that, modulo some fluctuations, 
the NMC steadily loses value against the 
BGU which like the BTC cannot possibly suffer from inflation generated by increased circulation. 

Now the
real money (that is BGUs) in control of some agent $P$ cannot be borrowed to $Q$. Borrowing (as well as
debt and interest for that reason)
makes no sense for an EXIM which only distinguishes between $P$ having access (control) or not. 
Instead, if $P$ intends to support $Q$ by means of a loan, that loan will be expressed in the unit of NMCoin, 
say a loan of  $l$ NMC.  
$P$ can borrow that amount (the principal sum $l$, not referred to as money but merely as a sum of near-money),
against a fee $f$ formerly called interest (expressed and redeemed in NMC) and return (redeem)
after the loan has expired, the principal back to $Q$ incremented by the fee, that is $l^{\prime} = l + f$. 

Payment of this fee may be considered unproblematic from the perspective of ethics based interest prohibition
because  the quantity $f$ NMC is not considered to constitute
a sum of money. When the loan is terminated after the agreed period,
$P$ can change the NMCs 
received from $Q$ (principal $r$ plus fees $f = l^{\prime} - l$) back to BGUs. 
This way of proceeding introduces for $P$ the risk that at this later stage $l+f$ BGU is insufficient to regain the 
original amount of $l$ BGU. Primarily the fact that $P$ is subject to the risk that $P$ expects in this second 
transaction constitutes a justification for 
$P$'s claim made on being paid fee $f$. This justification might be acceptable also
 from the perspective of opponents of interest payment. 

\subsection{A dual system may remedy ethical worries}
This viewpoint indicates a solution to the ethical issues raised in point~\ref{Getting-started} of Section~\ref{ethics}
below. 
By an adequate distribution of 
roles between an EXIM (say Bitguilder) , and one or more TINMs each need can be addressed.

A dual system provides much more operational flexibility than a system consisting of a single EXIM. 
Whether or not Bitguilder (being ``isomorphic'' to Bitcoin)  is sufficiently flexible to allow playing all roles of
a conventional money after combination with a suitable MTINM is a question that we cannot answer, moreover that
question belongs to theoretical 
economics rather than to informaticology.

\subsection{Mining rights: removing a single risk of failure}
If a Bitcoin successor, say BitguilderPlus plays a key role in an economy, the eco-footprint of mining must preferably
be reduced and too powerful mining cartels must be prevented. 
 The possibility that such dominance is established constitutes a ``single risk of
failure''. Here are some options for the reduction risks arising from mining:
\begin{itemize}
\item Instead of mining on miner owned or miner managed hardware, mining puzzles might be be 
solved by algorithms working for a 
fixed multi-core virtual machine. Now the puzzle is to develop by means of automatic programming 
fast algorithms for certain tasks. This approach would replace some of the computer work by human design work. 
Redesigning the mining competition for a lower eco-footprint is an interesting problem

\item A simpler option is to allow a setting with mining-rights to be distributed fairly over the world-population. 
Mining rights may be granted to agents who represent say 50,000,000 Citizens or multiples thereof, 
but no more than, say 250,000,000, in order to have a sufficient number of independent miners at any time. 
Smaller communities may also
participate in mining if they can collect the funds for acquiring suitable and competitive systems. Prevention of 
dominant mining coalitions seems to require some international law enforcement, however.

\end{itemize}

\section{Some remarks on Bitcoin ideology}
In this Section we will comment on three topics: (i) the experimental status of Bitcoin, (ii) paradoxical aspects of
Bitcoin as a software development project, and (iii) the Bitcoin business case.

\subsection{Bitcoin is experimental}
On a myriad of blogs and increasingly within conventional media Bitcoin is commented 
and criticized. Remarkably many commentators require Bitcoin to show a stability of value
against some mainstream conventional currencies that many conventional monies in 
economically weaker areas don't show.

There can be no doubt that Bitcoin is experimental. The following is unknown, and is being put to the
test with Bitcoin:
\begin{itemize}
\item Can a community of OSS programmers stay in the lead of the Bitcoin client
production process. The vagueness of the Bitcoin specification provides this community with a headstart,
but as soon as functional specifications for the Bitcoin client become standardized software 
components
there is no reason why an open source developer community would be in an advantageous position.
\item It is not known if a Bitcoin-like money can develop some stable value (as measured in 
some external form of purchasing power). Although it cannot be excluded that unmanaged automated
trust engineering creates stable values, it has not been proven either. This is an economic matter. 
If many transactions are carried out in Bitcoin, price levels may develop latency and stability 
without any form of governance.
\item Will the cryptography used in Bitcoin be sufficiently strong in the long run. And if not, what will
happen? (Has the set of cryptographic tools used in Bitcoin been chosen too weak on purpose so that 
the whole thing is like a Ponzi scheme after all, only to crash in say 2027 or so?)
\item Will we see the emergence of a ``secure Bitcoin standard'': featuring much stronger 
cryptography plus interface 
standards capable of strengthening cryptography if that needs to be done centuries from now (instead of 
so-called de facto implementation standards (which aren't real standards) that currently dominate the stage? 
\item If Bitcoin proves successful, will that induce the emergence of many similar competing systems or
not, and if so, what kind of competition is that going to be.
\end{itemize}

\subsection{Paradoxical aspects of Bitcoin}
Bitcoin provides several  remarkable paradoxes: 
	\begin{description} 
	\item[\em Expectation Paradox.] If Bitcoin survives until 2040 it must be very 
	strong by that time, and due to its design the BTC must then have risen to an 
	astronomic value, perhaps millions of Euro's.\footnote{%
	The WIKI on {\url www.Bitcoin.org} indicates that once sufficiently matured 
	the Bitcoin system might survive, possible at a reduced scale, the emergence of a
	technologically superior competitor on the market of cryptocurrencies.}%
	That development however, 
	viewed from the perspective of the agent maintaing a wallet with Bitcoin quantities, 
	is merely a deflationary trend. The trend is so strong, however, that who believes in 
	this future will become very reluctant to spend any BTCs. 
	Deflation makes the use of Bitcoin as a barter catalyst irrational.
		
	When not used for practical purposes Bitcoin will probably
	become irrelevant and inflation (in terms of 
	Bitcoin pricing of externalities) will predominate, making the practical use of 
	Bitcoin increasingly impossible. Thus if it will be heavily made use of,  
	Bitcoin will mainly be used, and for that reason be effectively created, 
	by those who don't believe in its ultimate victory over existing and 
	future centralized financial technologies.

	\item[\em Programmer Loyalty Paradox.] As an expression of the open source community Bitcoin
	collects loyalty from many software engineers who believe in open systems development.
	That view allows no obstruction or hindrance to be created when other engineers 
	(and may be former Bitcoin programmers who are unafraid of losing their own BTCs) 
	use the body of existing software in order to design
	a competing product. At the same time being involved in Bitcoin also generates loyalty with
	those who have secured themselves access to Bitcoin quantities now trusting the system 
	and using it as a store of value mechanism. These two loyalties are on the long run
	incompatible.\footnote{%
	Seemingly inconsistent loyalties can be an issue for contributors to all open 
	source efforts where software products
	are produced that have an independent economic value for their users which may be damaged for
	some uses if further development chooses priorities that deviate from those of these particular users.
	It seems that with Bitcoin the issue is more pronounced than with other projects and it is especially difficult
	to determine the value that can be destroyed when a new approach in software 
	architecture and development is chosen.}
	
	\item[\em Standardization Paradox.] Are Bitcoin OSS 
	programers committed to engineering products (official
	clients)  that lead the way to progress by introducing hardly noticeable 
	pragmatic modifications of the client behavior through  a chain of software releases, thus creating
	no more than a so-called de facto standards, with the ``protocol'' being an informal and dynamic
	entity on which nobody outside the developer team can rely ;
	or are programmers committed to the development of a level playing field for the entire 
	software engineering community by developing an open standard, that is a family of
	specifications of intended behavior for Bitcoin clients of various kinds, independently of 
	their implementation. It is hard to see how both ambitions can be reconciled.
	
	\item[\em Ideological Paradox.] At superficial inspection one might arrive at the conclusion that
	the open source software community has now come to the conclusion that a stable and truly
	international financial system can operate without a central bank and without 
	fractional reserve banking and without top level quantity management for 
	balancing between inflation and deflation.

	None of such conclusions go hand in hand with open source cryptocurrency development 
	as an ambition. That is:
	if the Bitcoin project happens to prove the opposite of these viewpoints its success as an
	open source approach to the future development of monies is just as great. Open 
	source software development must be unbiased (at least to a large extent)
	against the legal and economic aspects of managing money.\footnote{%
	If a project is started to engineer control programs for a pacemaker in OSS mode, then it
	is taken for granted that the OSS programmers involved in the project  
	would misuse their role when imposing self-invented heart frequencies
	on their prospective clients. Instead the programmers know that such inputs must come from
	a cardiologist. Similarly Bitcoin's pretenses as a money must not simply emerge from 
	computer programmers minds.}
	
	\item[\em Legality Paradox.] Bitcoin ``strives'' for official and normal prominence, so it seems, but
	its strength is really shown beyond any doubt if it survives a world wide prosecution. It may turn out to be at 
	least as strong as the internet, in no need of any official confirmation.\footnote{%
	In~\cite{Sova2013} it is considered unlikely that governments wile able to block the Bitcoin P2P network.}
	\end{description}
	
\subsection{Bitcoin business case.}
\label{ethics} 
The ideology, or simply business case, of Bitcoin is not easy to grasp. Active Bitcoin proponents sometimes
produce claims that are difficult to follow. Probably there is no such thing as a 
Bitcoin ideology. Different
participants may be involved in Bitcoin with very different objectives, with those who express idealistic
ambitions possibly in a minority. 

As a project of a group of programmers who upload their results on a repository for open source software,
the Bitcoin operation seems to be morally neutral at first sight. Can it be wrong to write programs and 
to give these away for free within the open source paradigm?
Probably yes, for instance if the programs constitute addictive betting 
schemes that drive their users into helpless
dependency.\footnote{%
Bitcoin creates a great opportunity to exploit botnets for mining (see~\cite{Plohmann2012}). 
This open invitation to computer abuse should probably not be held against the designers, programmers, and
users of Bitcoin. In~\cite{BregasB2012} a rather pessimistic perspective on the ethics of Bitcoin is put forward.} 
But Bitcoin can be approached in much more positive terms of course. 

Here we wish to criticize
some of the claimed advantages of Bitcoin, without claiming that these critical remarks suffice to arrive at some kind 
of judgement about Bitcoin as a movement, or  about the plausibility of anyone's own current, intended, 
or contemplated, involvement in it.
\begin{description}
\item[\em Open source advantage.]  It is claimed that as a financial system 
Bitcoin profits from the open source strategy. Here we have 
these objections:
	\begin{enumerate}
	\item As an open source endeavor, the production and distribution of open source ``official'' Bitcoin 
	clients is primarily a service to the user community of that software. 
	\item As a project or movement in financial systems engineering the merits of the 
	``Bitcoin movement''   would be much easier to understand 
	if the production process were carried out as an 
	open standards design project (with a Bitcoin client specification as proof of concept, which is 
	in competition with other
	client production lines for being a most widely used (de facto standard) implementation).
	\item There is no reason why the production or the use of commercial and closed source Bitcoin clients
	is against the spirit of Bitcoin. Bitcoin is P2P, software only, and cryptography based. But it is not open source based
	as the open source software engineering process cannot be essential for it. To claim the expectation that open 
	source software engineering for Bitcoin clients outperforms closed source software engineering for Bitcoin clients is
	as far as one can get. 
	\item Given the open source nature  of the 
	Bitcoin effort (and in the absence of enforceable specifications) there seems to be no defense against 
	an ``attack'' where alternative clients are intentionally programmed in order to arrive at a hard fork which can be misused
	by the designers of the alternative programs for their own (malicious or merely non-cooperative) purposes.
	\end{enumerate}
	
\item[\em Bitcoin failing on anonymity.]
Bitcoin is said to facilitate anonymous operation, (which in addition is supposed to be a justified 
objective, at least in principle).

But it is well-known that making use of Bitcoin transactions in an anonymous way is not an easy task.\footnote{%
In~\cite{AndroulakiKRSC2012} one finds an experiment with Bitcoin usage in the ETH Z\"{u}rich from which the 
conclusion is drawn that daily use in a University setting exposes some 40\% of the users to a breach of anonymity. 
See also~\cite{ReidH2013}.}
Doing so requires an intimate awareness of existing technique for anonymity engineering often
predating Bitcoin.\footnote{%
One might hold that Bitcoin allows for more transparency and less anonymity than conventional
currencies do. Two principals handing over a quantity of $q$ BTC can always prove afterwards 
that the transfer has 
taken place by proving their control over the private keys involved. A corresponding transparency is absent 
in the setting of conventional cash transactions.}

\item[\em Mining Bitcoins: just for the happy few.]
 It is sometimes claimed as an advantage that everyone can start constructing Bitcoins for 
his or her own use through mining. 

That may be true in principle though it suggests a deceptively simple picture as mining
has already become a very competitive and costly activity. It also hard to reconcile
with the often repeated assertion that transaction costs in the Bitcoin system are low. Increasingly
miners need to acquire BTCs by way of transaction fees provided by participants performing transactions.

\item[\em Bitcoin invisibly controlled by  a centralized authority.]
Indeed the quantity of circulation in Bitcoin is not controlled by a central authority but everything else seems
to be. The Bitcoin community has its own way to create and maintain a power hierarchy which has been successfully
developed through many years within the open source community. Technical problems in the Bitcoin software
need to be solved and without decision taking that may impact on all participants some problems cannot be solved. In
\cite{Grinberg2012} the improper use of discretionary authority (leading to public loss of confidence) 
by the Bitcoin developer community is mentioned as a potential risk factor.

In~\cite{Plassaras2013} it is outlined how Bitcoin might come to pose a threat for the international financial
system, and how and against which cost the IMF may be able to offer protection against such threats.
\end{description}

\section{Conclusions}
Informational monies have come to stay and will be a topic of much future research. 

Money can by way of comparison be understood in terms of publishing, (including copyrights, distribution, circulation,
and ownership).
It can also be understood in terms of information and information processing. The internet provides a platform for
mounting a variety of attacks on the status of key players in the classical paradigms of publishing style money which will be 
comparable to the attacks that were developed on the classical forms of publishing in other areas. These attacks
are now ongoing, but giving an assessment of their strength is still a matter of speculation. 
 
With full appreciation of past performance one may state that conventional views of
money should not be given a definitional status for the concept of money. That concept is under development,
and the concept of money it is far too important for the development of its meaning to be left to the discretion of
politicians,  economists, lawyers, and and bankers only.

Bitcoin might prove to be the paradigmatic development in the field. Like Schr\"{o}dinger's cat may be 
simultaneously death and alive, Bitcoin may
simultaneously be a TIM and an EXIM, with a high probability of being seen as a TIM when observed from a court.
This remarkable superposition of ideologies helps to make Bitcoin both conceptually powerful and technically attractive. 
 
 {\bf Acknowledgement.} Discussions with Andrea Haker, Alban Ponse, and 
 Sanne Nolst Trenit\'{e}, (all University of Amsterdam), as well as comments made by Wan Fokkink 
 (VU University Amsterdam) on \texttt{v1} of the paper have contributed to this work. The first author 
 is indebted to his daughter Machteld Bergstra for critical remarks about a draft of the first version of the paper.
 Of course the authors are fully and only responsible for what has been written.
\bibliographystyle{plain}

\appendix
\section{Promises}
The use of Bitguilder in practice is embedded in a network of promises which are delivered in a setting of mutual trust.
Therefore it matters to have some theory of promises available that connects well with the 
setting of P2P networks and EXIMs.

In~\cite{Burgess2005,Burgess2007,BergstraB2008} proposals have been made about a theory of promises 
intended for application in distributed computing. An
important claim made in these papers is that promises made by 
agent (promiser) $P$  will not give rise to obligations that 
are binding for $P$. Further a promise is made by a promiser $P$ is 
made to a promisee (or to a group of promisees), say $Q$ (or $G$), and it has a promise
body containing information about what $P$ must see to pertain, to 
have done, or to achieve in order to live up to the promise,
and finally there must be an  
audience (also called scope) $S$ which comprises a collection of 
agents that contains $P$ and that may or may not contain $Q$ (or some or all members of $G$).

From a philosophical point of view every promise made by $P$ gives rise to a promissory 
obligation for $Q$, which may
in some cases be vacuous, a flexibility given with (and needed for) the notion of a 
promissory obligation. Therefore rejecting
the obligation generating aspect of promises seems to be futile. However, if one insists that in some setting
promises and obligations have bodies that are of the same type, the matter changes and the setting of 
Bitcoin-like EXIM, say Bitguilder with unit BGU provides an example of just that situation.

A typical and unproblematic promise made by promiser $P$ to promisee $Q$
 is that before time $t$ has expired $P$ will see to it that $p$ BGU is 
transferred from  public key (that is account) \texttt{pk$_1$} to public key \texttt{pk$_2$}.

Here it is assumed that $P$ controls \texttt{pk$_1$} and $Q$ controls \texttt{pk$_2$}. 
The audience may include several other
agents. 

The observation that can be made at this point
 that there cannot exist an obligation for $P$ to make that same transfer, because the existence of
such an obligation contradicts the autonomy of $P$ which is assumed for any EXIM, including
Bitguilder. In this setting the 
assumption that promises
have no complementary obligations expresses the idea of autonomy for Bitguilder participants. 
An obligation not to make a certain
transfer cannot exist either for the same reason. Thus in the world of Bitguilder transfers 
(and for Bitcoin transfers just as well) there is
ample room for promising about transfers, and for complementary counter promises 
and their timing but there is no room for obligations of a similar kind.

A simple functionality of a promise with nontrivial scope (scope different from promisee)
 is as follows: 
\begin{enumerate} 
\item $P$  promises
to group $G$  including $Q$, with scope $G$ to 
provide to at most $k$ agents satisfying condition $\gamma$ the service
$s$ in return for a payment $p$ BGU.
\item  $Q$ promises to all 
members of $G$ (with scope $G$)
 that requests will be served by $P$ in the order of arrival,
\item  $Q$ promises to $G$ with scope $G$ that BGU transfers 
can be made into \texttt{pk$_1$}. Then after taking notice of $P$'s 
successive promises, 
\item $Q$ promises to $P$ with scope $P$ that it satisfies $\gamma$, and 
\item $Q$ promises to $P$ with scope $P$  that it will transfer 
$p$ BGU from \texttt{pk$_2$}  to \texttt{pk$_1$} before $t$ has elapsed. Finally, 
\item $P$ promises to $G$ with scope $G$ that service $s$ will be provided once
a transfer of $p$ BGU has been received from \texttt{pk$_2$} to the sender of that transfer
and under the condition that this sender indeed satisfies $\gamma$.
\end{enumerate}

After all these preparations $Q$ can autonomously effectuate the transfer of $p$ BGU from
\texttt{pk$_2$}  to \texttt{pk$_1$}. The virtue of this sequence of promises has been (i) that $P$ knows 
that the transfer is made and (ii) that it will now respond by delivering $s$ to $Q$ (provided
$\gamma$ is satisfied by $Q$ in conformance with $Q$'s promise (no. 4). In addition members of
$G$ know that a service option has been allocated to $P$. To the extent that it is unknown to members of $G$
who pays via \texttt{pk$_2$} the anonymity set of that public key (to which $P$ may prefer not 
to be openly linked) has been reduced only from an original $A$ to $A \cap G$. At the same time
$Q$ can work out an expectation that service $s$ will be delivered, taking account the trust
$Q$ has developed in $P$ as well as $P$'s reputation in $G$.

The final promise is plausible if (i) $P$ knows that $Q$ prefers to remain anonymous, and (ii)
the anonymity set of \texttt{pk$_1$} is $G$, and (iii) $P$ intends to proceed (by way of making promises)
in such a way that members of 
$G$ can monitor progress by means of block chain inspection. There are several alternatives for the final promise:
\begin{enumerate} 
\item (if anonymity of $Q$ and the account it will use does not matter)
$P$ promises to $Q$ with scope $G$ that service $s$ will be provided once
a transfer of $p$ BGU has been received from \texttt{pk$_2$} to the sender of that transfer
and under the condition that this sender indeed satisfies $\gamma$, or

\item (if it does not matter that members of $G$ can see that the transfer has been made by
inspecting the block chain and if $P$ knows that 
$Q$ prefers to remain anonymous) $P$ promises to $G$ with scope $G$ that service $s$ will be provided once
a transfer of $p$ BGU has been received from a non-disclosed account to the sender of 
that transfer
and under the condition that this sender indeed satisfies $\gamma$, or
\item (if anonymity of $Q$ is immaterial, but $P$ knows that $Q$ prefers not to be publicly linked to the
account that it will use\footnote{%
In spite of that account having the status of a public key.}) $P$ promises 
to $Q$ with scope $G$ that service $s$ will be provided once
a transfer of $p$ BGU has been received and under the condition that $Q$
 indeed satisfies $\gamma$.
\end{enumerate}

In this example promises are made in order to modify the expectations of other agents, and to prepare
other agents so as to prevent problematic surprises. For instance if $Q$ makes the transfer after 
receiving $P$'s promises, then $Q$ may not even know to whom the service is to be provided. If $Q$ also 
sends a message to $P$ asserting that it has transferred $p$ BGU from account  \texttt{pk$_2$} with
the expectation that it will, be served $s$, then $Q$ may run out of options if that takes place
more than $k$ times. After exchanging an alternation of conditional promises an agreement is made
with some public visibility while preferred characteristics of anonymity can hopefully be preserved.

\section{Units for monies of account}
\label{BGnotation}
In \cite{Bergstra2013a} formalbitcoins (FBTCs) have been put forward as units meant for use in texts 
about the circulation of Bitcoin in operation. FBTC is not meant as a unit of account, rather it is meant to provide a tool for hypothetical 
reasoning about BTCs. Units of account may be viewed as yet another type that comes with a unit of a (near-)money.
We will discuss units of account in the context of Bitguilder thus avoiding a corresponding 
development for Bitcoin at this stage
because of its experimental character.

Money of account refers to an other use of money than exchange and store of value. 
It is commonly taken for granted that such can be done with the same notational conventions as are used when dealing
with exchange and storage  of amounts.

We consider it promising to experiment with the use of different but related units for account.
We will illustrate that in the case of Bitguilder by having a specific notation for quantities of 
Bitguilder units which are meant to
indicate measurements, estimates, prices, and other quantities that do not refer to existing volumes of 
Bitguilder informational coins.

Together with the Bitguilder informational coin, which is an exclusively informational coin that we will call 
BGU for Bitguilder unit, there might 
be a corresponding unit of valuation, called BGUA for Bitguilder unit of account. 
In principle one expects that a valuation function $V_{\texttt{bg}}$ is available that satisfies the following equation:
\[V_{\texttt{bg}}(q\,\, \texttt{BGU}) = q\,\, \texttt{BGUA}.\]

A value can be negative if it refers to an obligation one would be happy to be relieved from through a 
payment in BGU, and it can be arbitrarily small even in the absence of a micro payments system that 
permits transfers of such small quantities. 

Similarly a value can be higher than the total number of BGUs that have been or will be created within the system. 
For instance in the sentence that an observer of Bitguilder might produce at some moment in time: 
``Bitguilder can be improved if the maximum of circulating BGUs is increased to a value of $q$ BGUA''.

As an example let $\texttt{Access}_{P}(a,t)$ be a boolean that expresses that user $P$ has access to public key $a$
at time $t$, let $q(a,t)$ denote the dimensionless amount of BGU that is accessible via $a$ at time $t$ for an
agent availing of the corresponding secret key, 
and let $V_{\texttt{bitguilder}}(a,t)$ represent the BGUA measured value attached to $a$. Then the Bitguilder 
based wealth of $P$ can be defined as follows:
\[ \texttt{BGU1}_{\texttt{wealth}}(P,t) =  \sum_{a \in \texttt{pks}(t) \& \texttt{Access}_{P}(a,t)}  
\frac{q(a,t)}{\#_{Q}(\texttt{Access}_{Q}(a,t))}\,\,\texttt{BGUA.}\]
Here $\#_{Q}(\texttt{Access}_{Q}(a,t))$ counts the number of users who have access to $a$ at time $t$. A 
difficulty with this definition is that it depends on ``having access to $a$'', a far from trivial notion for which a 
formal definition is missing.
\subsection{Formal BGUs: FBGUs}
When writing about Bitguilder one may want to speak of hypothetical (formal) BGUs that circulate in a model of Bitguilder
rather than in Bitguilder itself. That leads to BGU terms (textual indications of BGU quantities) which don't refer
to any actual BGU quantity in the real system and also don't refer to any valuation. 

We consider it justified to extend the typing of informational units and derived concepts to the extreme and to speak
formally of FBGUs rather than to be implicit about the fact that one is modeling a system of BGU 
circulation or a potential further development of the Bitguilder blockchain.\footnote{%
This suggestion follows a proposal concerning the use of FormalEuros in \cite{Bergstra2013a}.}

As an example consider the following equation that might be used when experimenting with a design for 
an extension of the Bitguilder system which incorporates planning for regular payments.

$P$ promises an external agent $B$ that each year $y$ at a fixed date $y+t$, $P$ transfers an amount $a$ 
by way of taxation:
\[ a = \texttt{BGU}_{\texttt{taxation}}(P,y+t) =  \frac{1}{100} \cdot \sum_{a \in \texttt{pks}(y+t) \& \texttt{Access}_{P}(a,y+t)}  
\frac{q(a,y+t)}{\#_{Q}(\texttt{Access}_{Q}(a,y+t))}\,\,\texttt{FBGU}\]
to an account named that year by $B$.
The F in FBGU indicates that one does not expect any true transactions to be performed on the basis of this
promise. A separate action will be needed for $P$ to turn the promise into a transfer. 
\footnote{%
The justification for working with FGBU rather than with BGUA
becomes  even stronger if the promise is contemplated as a part of the design a payments scheme involving several 
other yet uncommitted series of transfers.}

We find that
\[ \texttt{BGU}_{\texttt{taxation}}(P,y+t) =  \frac{1}{100} \cdot  \texttt{BGU1}_{\texttt{wealth}}(P,y+t)\,\,\frac{\texttt{FBGU}}{\texttt{BGUA}}.\]

\subsection{Formal BGUAs: FBGUAs}
When the economy of a hypothetical instance of Bitguilder (or BitguilderPlus) is investigated then the need to formalize the
corresponding unit of value arises as well. That leads to FBGUAs.

A difficulty of the two examples just mentioned is that Suppose that $\texttt{Access}_{P}(a,t)$ cannot be
computed automatically. In order to find a definition with more operational meaning an option is to
one replace $\texttt{Access}_{P}(a,t)$ with a practically  implementable version of it, for instance the following:

$\texttt{cAccess}_{P}(a,t)$ (for confirmed access), which is inductively defined as follows: $\texttt{cAccess}_{P}(a,t)$
holds if either:
\begin{enumerate}
\item $P$ has confirmed upon request by an external
agent $E$ that $P$ has had access to $a$ at some moment before $t$, and $P$ has in addition demonstrated access  to
 $a$ also before $t$, or 
 \item some other agent $Q$  for whom  $\texttt{cAccess}_{Q}(a,t)$ has 
 been established already,   has asserted upon request by $E$ that it shared access 
 to $a$ with $P$ at some moment $r$ before $t$.
  \end{enumerate}

 We do not claim that this is a valid definition of having access to $a$, for instance $P$ may deliberately not have confirmed
 its access to $a$ upon $E$'s request (false negative), or $Q$ 
 may lie about having (had) shared access with $P$ (false positive), or $E$ may have forgotten to ask $Q$ about it 
 (false negative), or $B$ may have lied about having seen $P$'s proof of access to $a$ (false positive),
 and so on, but in any case it gives some indication concerning how a 
 definition that can be used in  practice might look like.
 
The ``new" definition of accessibility of an account for an agent $P$ can be used as a building block in a formal definition of  
wealth for $P$. Being a non-asset rather than an indication of an existing sum of money it is measured in 
formal units of account. $\texttt{BGU2}_{\texttt{wealth}}(P,t)$ as an alternative for the definition of $\texttt{BGU}_{\texttt{wealth}}(P,t)$.
\[ \texttt{BGU2}_{\texttt{wealth}}(P,t) =  \sum_{a \in \texttt{pks}(t) \& \texttt{cAccess}_{P}(a,t)}  
\frac{q(a,t)}{\#_{Q}(\texttt{cAccess}_{Q}(a,t))}\,\, \texttt{FBGUA.}\]

A justification for the use of FBGUA as a unit may lie in the awareness that one is dealing with a stage of design of
financial schemes and practices from which proper accounting methods can only be extracted in a later stage, so that
at present FGBUA can range over possible forthcoming realizations of BGUA.\footnote{%
Distinguishing BGU and FBGU occurs if one intends to speak of hypothetical quantities that might exist in another
world though with the same system of payment. Following~\cite{Bergstra2013a}  FormalCoins for Euro would be used if
on speaks of Euro coins when designing a vending machine. A new vending machine design 
is comparable to a parametrized data type with a variety of coin types a formal parameters. After a realization 
of the design has be completion and is put into operation (perhaps during a tea), instead of the formal parameter
a real coin is passed as an actual parameter.}
\subsection{Complementary units for BTC}
Similar to what we have written on Bitguilder above, besides the BTC a unit of account BTCA can be assumed, and besides
the FBTC a unit FBTCA can be assumed for formal (in the sense of meta) accounting purposes.

\subsection{Some reflection about units and dimensions}
In~\cite{Bergstra2013a} an account is given of dimensions and units for monies. It was noticed that if EUR is used as the 
unit of Eurozone money, and if EUR is viewed as a dimension the for instance a dimension $\frac{1}{\texttt{EUR}^2}$ 
makes sense in some cases. It can be used to measure the productivity of efforts made by a consultant for 
increasing productivity of a production process.

We will contemplate the hypothetical situation in which Bitcoin has become the only 
money in existence after all other monies have faded away. We assume sustainable technologies for all processes,
so that cost in BTC are real and integral costs. Let us first consider some business cases for mining in different.
\begin{enumerate}
\item Assuming that miner $M$ has equipment $E$ at hand at operating cost $C_E$ per unit of time $U$ (that is 
$C_E \cdot \frac{\texttt{BTC}}{\texttt{U}}$). And assume that
$M$ expects to produce $e \cdot \frac{\texttt{BTC}}{\texttt{U}}$  (Bitcoin for mining at that stage are mostly 
earned by cashing fees), then the productivity of mining may be written as a dimensionless quantity
$\frac{ e}{C_E} $ while the business case for mining operations performed by $M$ in this way simply 
requires that $C_E \leq e$.
\item Assuming that:
\begin{enumerate}
\item new and improved equipment is available at cost $C_{E^{\prime}}$, 
\item that all equipment (new or old) is written off in $t$ units of time,
\item that using the new equipment productivity becomes
$e^{\prime} \cdot \frac{\texttt{BTC}}{\texttt{U}}$, 
\item that old equipment cannot be sold and must be returned to its manufacturer for free 
(in order for $M$ to be entitled to new equipment), 
and
\item that old equipment needs to be replaced in any case but that it can be replaced by equivalent but new equipment 
at the same cost as before. 
\end{enumerate}
Then the business case requirement for replacement of equipment by the same equipment is
$t \cdot e \geq  C_E$ and the business case for replacement by
 improved equipment requires: $t \cdot (e^{\prime}- e) \geq  C_{E^{\prime}}-C_E$.
 
 \item If the proportion of all property (all property expressed in BTCA) to the BTC basis 21 million remains constant
 then BTC must be an EXIM at least in the sense that no agent can be safeguarded against accidental loss of BTC's (doing so requires handing over other wealth, but more problematically it increases the valuation of all property because the safeguarding
 entity must buy ``other wealth" to compensate the losing agent for its loss.
 
 \item In a ``Bitguilder only" world all agents need very good physical security for their secret keys. 
 They simply cannot recover from the loss of one or more keys and that may well be the consequence of wars, earthquakes,
 tsunami's etc.
 
 \item It seems that a Bitguilder only world is unpleasant because it exposes agents to very high risks, 
 while a Bitcoin only world (where ownership based insurance policies are used to protect agents against 
 not-selfinflicted Bitcoin loss) deflates with the predictably non-zero rate of Bitcoin loss, which is also quite unattractive
 on the long run.
\end{enumerate}

\subsection{Syntax for moneys of account}
Given a money of account and a unit of value for that money, the main application of it is to allow writing texts and technical
documents with quantitative information about money. That may include specific budgets, budget models, 
parametrized prices, and accounting models.

Rational number expressions with or without free variables are essential for the syntax to be employed. 
What we intend to point 
out in this Section is that the syntax of rational numbers can be approached in the following manner.
\begin{enumerate}
\item A formal syntax is needed with operation symbols for all constants and functions, including division. 

\item Mathematicians usually don't accept the concept of a function symbol and 
once division appears on the scene, logicians try to avoid introducing a  formal symbol
for division because that is not strictly needed if first order definitions are used. It appears that conventional 
mathematical style avoids to give any prominence to the notion of syntax 
while logical treatments  preferably avoid partial functions although more often than not only implicitly.
\item Once division is a function symbol in the syntax for rational numbers the question about the value of $1/0$ cannot
be avoided and a systematic treatment of financial figures will require some decision on that matter.
\item Several options exist to deal with the meaning of $1/0$, and we are in favor of the notation of meadows 
(see~\cite{BBP2013}, \cite{BM2011a}, and further work cited there) as a candidate
for the simplest solution for these matters.
\end{enumerate}

\subsection{Virtual monies based on Bitguilder}
In \cite{Bergstra2013a} the notion of a virtual money is described in correspondence with classical computer science
concepts of virtual machine, virtual memory, and virtual network. That is a virtual money is expressed in terms of another ``base money'' and it appears to its users as another money whereas in fact its handling consists of sequences of 
interactions in terms of an underlying money. Neither BGVU nor FBGU are meant to be units in a virtual money because
both occur in systems of account but not in transactions.

Virtual monies can be based on Bitguilder, or like in \cite{Bergstra2013a} can be extracted from patterns of (hypothetical) 
use of BGUs. Given the hypothetical status of Bitguilder, contemplating such virtual monies is rather remote. 
However, contemplating virtual monies on top of Bitcoin is much less remote.
\section{Survey of Bitcoin participant roles}
Bitcoin induces a steadily growing hierarchy of participants. 
Here is a survey of different roles that have emerged thus far:
\begin{enumerate}
\item Developers: most prominently there is a developer team 
which is firmly in control of the design, production, distribution, and maintenance,
of a particular brand of Bitcoin software clients which is often understood, portrayed,
or referred to,  as being official or preferred. Other Bitcoin clients exist, however.
\item  Users: users are currently differentiating in
	\begin{itemize}
	\item  ordinary users: these are running their private client under their own and sole 
	responsibility. Ordinary users profit from Bitcoin as a means of exchange and as a store of value.
	\item Bitcoin (participation) service providers: these provide several service types, such as:
		\begin{itemize}
		\item exchanging Bitcoin for quantities non-Bitcoin monies (e.g. the Euro), 
		\item managing wallets (containing informational coins),
		\item managing transactions between participants, and
		\item providing chartal market information for comparing the BTC with units of mone
		outside the Bitcoin system.
		\end{itemize}
	\item customers of Bitcoin service providers: customers delegate the task of operating a Bitcoin 
	client to their service provider (which can be done in different ways). 
	These customers simulate
	being a peer in the Bitcoin P2P network without ever contributing to it otherwise 
	than by offering fees for miners who successfully validate their incoming 
	and outgoing transactions.
	\end{itemize}
\item Miners: miners split at least in three groups. We mention proper miners, (honest) poolmasters, and (dishonest) 
botmasters). Miners take the responsibility to maintain the integrity of the historical log of 
transactions (the blockchain), and miners produce new or existing 
coins as a reward for that activity. Miners are  by definition 
users as well though they may be hoarding their earnings indefinitely, at least in principle.
\item Peripheral agents: a final  category of participants contains all peripheral  agents. Here is a non-exhaustive listing
of their roles:
	\begin{itemize}
	\item  producers and sellers of ASICs and other dedicated equipment and software  for mining, 
	\item hosts of websites and blogs about  Bitcoin, 
	\item market analysts,
	\item legal experts dealing with authorities taking an interest in Bitcoin, 
	\item legal experts representing legal or financial authorities in defining a position with 
	respect to questions posed by Bitcoin's existence, and
	\item researchers and academics who investigate the foundations of Bitcoin, ranging from its
	cryptography to its economics and sociology, and possibly
	\item participants with deviant objectives. For instance the Bitcoin system provides a 
	very extensive testbed for the cryptographic operation SHA--256.\footnote{%
	If that hashing function can ever be broken, it may 
	well happen as a consequence of its massive use in the proof of work component of 
	Bitcoin mining. One 
	might even hold that Bitcoin merely constitutes a multiplayer 
	game (an applied game or serious game in fact) that has been designed
	with the primary objective to successfully
	cryptanalyze SHA--256. As long as the identity of the Bitcoin designer remains unknown
	it is hard to confirm that his/her/their stated (in~\cite{Nakamoto2008}) objectives coincide with 
	``true" objectives.}
	\end{itemize}
\end{enumerate}	

\end{document}